\documentclass[12pt]{article}
\usepackage{amsfonts,amssymb,amsmath}
\usepackage{graphicx}
\usepackage{float}
\usepackage{xcolor}

\newcommand{\RR}{\mathbb R}
\newcommand{\CC}{\mathbb C}

\newcommand{\ZZ}{{\mathbb Z}}

\newcommand{\LG}{{\mathfrak L}}
\newcommand{\Den}{\mathrm{Den}}

\renewcommand{\Re}{\mathop{\mathrm{Re}}}
\renewcommand{\Im}{\mathop{\mathrm{Im}}}

\newcommand{\tr}{\mathop{\mathrm{tr}}}

\newcommand{\beq}{\begin{equation}}
\newcommand{\eeq}{\end{equation}}
\newcommand{\ba}{\begin{array}}
\newcommand{\ea}{\end{array}}
\newcommand{\bea}{\begin{eqnarray}}

\newcommand{\eea}{\end{eqnarray}}
\newcommand{\eps}{{\epsilon}}

\begin{document}
\begin{center}

{\bf The effect of a small loss or gain \\ in the periodic NLS anomalous wave dynamics. I }
\vskip 10pt
{\it F. Coppini $^{1,4}$, P. G. Grinevich $^{2,5}$, and P. M. Santini $^{3,6,7}$}

\vskip 10pt

{\it

$^1$ PhD Program in Physics, Dipartimento di Fisica, Universit\`a di Roma ``La Sapienza'', and 
Istituto Nazionale di Fisica Nucleare (INFN), Sezione di Roma, 
Piazz.le Aldo Moro 2, I-00185 Roma, Italy
 
\smallskip

$^2$ Steklov Mathematical Institute of Russian Academy of Sciences, 8 Gubkina St., Moscow, 199911, Russia,  and 
 L.D. Landau Institute for Theoretical Physics, pr. Akademika Semenova 1a,
Chernogolovka, 142432, Russia 

\smallskip

$^3$ Dipartimento di Fisica, Universit\`a di Roma ``La Sapienza'', and \\
Istituto Nazionale di Fisica Nucleare (INFN), Sezione di Roma, \\
Piazz.le Aldo Moro 2, I-00185 Roma, Italy}

\smallskip

\vskip 10pt

$^{4}$e-mail: {\tt francesco.coppini@uniroma1.it}\\
$^{5}$e-mail: {\tt pgg@landau.ac.ru}\\
$^{6}$e-mail: {\tt paolo.santini@roma1.infn.it}\\
$^{7}$e-mail: {\tt paolomaria.santini@uniroma1.it}
\vskip 10pt

{\today}

\end{center}

\begin{abstract}
  The focusing Nonlinear Schr\"odinger (NLS) equation is the simplest universal model describing the modulation instability (MI) of quasi monochromatic waves in weakly nonlinear media, and MI is considered the main physical mechanism for the appearence of anomalous (rogue) waves (AWs) in nature. Using the finite gap method, two of us (PGG and PMS) have recently solved, to leading order and in terms of elementary functions of the initial data, the NLS Cauchy problem for generic periodic initial perturbations of the unstable background solution of NLS (what we call the Cauchy problem of the AWs), in the case of a finite number of unstable modes. In this paper, concentrating on the simplest case of a single unstable mode, we study the periodic Cauchy problem of the AWs for the NLS equation perturbed by a linear loss or gain term. Using the finite gap method and the theory of perturbations of soliton PDEs, we construct the proper analytic model describing quantitatively how the solution evolves, after a suitable transient, into slowly varying lower dimensional patterns (attractors) in the $(x,t)$ plane, characterized by $\Delta X=L/2$ in the case of loss, and by $\Delta X=0$ in the case of gain, where $\Delta X$ is the $x$-shift of the position of the AW during the recurrence, and $L$ is the period. This process is described, to leading order, in terms of elementary functions of the initial data. Since dissipation can hardly be avoided in all natural phenomena involving AWs, and since a small dissipation induces $O(1)$ effects on the periodic AW dynamics, generating the slowly varying loss/gain attractors analytically described in this paper, we expect that these attractors, together with their generalizations corresponding to more unstable modes, will play a basic role in the theory of periodic AWs in nature.     
\end{abstract}

\section{Introduction}

The self-focusing Nonlinear Schr\"odinger (NLS) equation
\beq\label{NLS_foc_defoc}
i u_t +u_{xx}+2 |u|^2 u=0, \ \ u=u(x,t)\in\CC, 
\eeq
is the simplest universal model in the description of the propagation of a quasi monochromatic wave in a weakly nonlinear medium; in particular, it is relevant in water waves \cite{Zakharov,AS}, in nonlinear optics \cite{Solli,Bortolozzo,PMContiADelRe}, in Langmuir waves in a plasma \cite{Malomed}, and in the theory of Bose-Einstein condensates \cite{Bludov,Pita}. Its homogeneous solution
\beq\label{back1}
u_0(x,t)=a\exp(2i|a|^2 t), \ \ \mbox{$a$ complex constant parameter,}
\eeq
describing Stokes waves \cite{Stokes} in a water wave context, a state of constant light intensity in nonlinear optics, and a state of constant boson density in a Bose-Einstein condensate, is unstable under the perturbation of waves with sufficiently large wave length \cite{Talanov,Lighthill,BF,Zakharov,ZakharovOstro,Taniuti,Salasnich}, and this modulation instability (MI) is considered as the main cause for the formation of anomalous (rogue, extreme, freak) waves (AWs) in nature \cite{HendersonPeregrine,Dysthe,Osborne,KharifPeli1,KharifPeli2,Onorato2}. 

The integrable nature of the focusing NLS \cite{ZakharovShabat} allows one to construct a large zoo of exact solutions, corresponding to perturbations of the background, by degenerating finite-gap solutions \cite{ItsRybinSall,BBEIM,Krichever2,Krichever3}, when the spectral curve becomes rational, or using classical Darboux transformations \cite{Matveev0}, dressing techniques \cite{ZakharovShabatdress,ZakharovManakov,ZakharovMikha}, and the Hirota method \cite{Hirota0,Hirota}. Among these basic solutions, we mention the Peregrine soliton \cite{Peregrine}, rationally localized in $x$ and $t$ over the background (\ref{back1}), the so-called Kuznetsov \cite{Kuznetsov} - Kawata - Inoue \cite{KI} - Ma \cite{Ma} soliton, exponentially localized in space over the background and periodic in time, the solution found by Akhmediev, Eleonskii and Kulagin in \cite{Akhmed0}, periodic in $x$ and exponentially localized in time over the background (\ref{back1}), known in the literature as the Akhmediev breather (AB), its elliptic generalizations \cite{Akhmed1,Akhmed2}, and its  multi-soliton generalizations \cite{ItsRybinSall}.  Generalizations of these solutions to the case of integrable multicomponent NLS equations, characterized by a richer spectral theory, have also been found \cite{BDegaCW,DegaLomb0,DegaLomb,DegaLombSommacal,DegaLombSommacal2}.

Concerning the NLS Cauchy problems in which the initial condition consists of a perturbation of the exact background (\ref{back1}), what we call the Cauchy problem of the AWs, if such a perturbation is localized, then slowly modulated periodic oscillations described by the elliptic solution of (\ref{NLS_foc_defoc}) play a relevant role in the longtime regime \cite{Biondini1,Biondini2}. The relevance of the Kuznetsov -- Kawata - Inoue -- Ma solitons and of the superregular solitons (constructed by Zakharov and Gelash \cite{ZakharovGelash1}, see also \cite{ZakharovGelash2}, \cite{ZakharovGelash3}) in this problem was investigated in \cite{Gel}.

If the initial perturbation is $x$-periodic, numerical and real experiments indicate that the solutions of NLS exhibit instead time recurrence \cite{Yuen1,Yuen2,Yuen3,Akhmed3,Simaeys,Amin,trillo,PieranFZMAGSCDR}, as well as numerically induced chaos \cite{AblowHerbst,AblowSchobHerbst,AblowHHShober}, in which the almost homoclinic solutions of Akhmediev type seem to play a relevant role \cite{Ercolani,FL,CaliniEMcShober,CaliniShober1,CaliniShober2}. There are reports of experiments in which the Peregrine and the Akhmediev solitons were observed \cite{CHA_observP,KFFMDGA_observP,Yuen3,Tulin,Amin,trillo,PieranFZMAGSCDR}. Their relevance within some classes of localized initial data for NLS, in the small dispersion regime, was shown in \cite{BT,EKT}; see also \cite{GT,TBETB} for the investigation of their relevance in ocean waves and fiber optics.

Using the finite-gap method \cite{Novikov,Dubrovin,ItsMatveev,Lax,MKVM,Krichever} (see \cite{ItsKotlj} for its first application to NLS), two of us (PGG and PMS) have recently solved \cite{GS1,GS2}, to leading order and in terms of elementary functions, the Cauchy problem of the AWs for the self-focusing NLS equation
\begin{equation}
\label{eq:nls1}
i u_t+u_{xx}+2|u|^2u=0, \qquad
u=u(x,t)\in\mathbb{C},
\end{equation}  
for a generic order $\epsilon$ periodic initial perturbation of the unstable background solution (\ref{back1}),
in the case of a finite number of unstable modes. Namely, the following Cauchy problem was solved:
\begin{equation}
\label{eq:nls_cauchy1}
u(x,0)=a\left(1+\varepsilon v(x)\right), \qquad
0<\varepsilon\ll 1, \quad
v(x+L)=v(x),
\end{equation}
where
\begin{equation}
\label{eq:nls_cauchy2}
v(x)=\sum_{j=1}^{\infty}(c_j e^{i k_j x}+c_{-j}e^{-i k_j x}),\qquad
k_j=\frac{2\pi}{L}j,
\end{equation}
the average of the initial perturbation is assumed, without loss of generality, to be zero, and the period $L$ is assumed to be generic ($L/\pi$ is not an integer). See also \cite{GS3} for an alternative approach to the study of the AW recurrence, in the case of one unstable mode, based on matched asymptotic expansions; see \cite{GS4} for the study of the numerical instabilities of the AB and of a finite-gap model describing them; see \cite{GS5} for the analytic study of the phase resonances in the AW recurrence; see \cite{San} and \cite{Coppini} for the analytic study of the AW recurrence in other NLS type models: respectively the PT-symmetric NLS equation \cite{AM1} and the Ablowitz-Ladik model \cite{AL}.  

It is well-know that the mode $k$ is linearly unstable if $|k|<2|a|$, implying that the number $N$ of unstable modes for the problem (\ref{eq:nls1}), (\ref{eq:nls_cauchy1}) is finite and given by 
\begin{equation}
\label{eq:nls_N}
N=\biggl\lfloor \frac{|a| L}{\pi}\biggr\rfloor,
\end{equation}
where ~$\lfloor x \rfloor$, $x\in\mathbb{R}$, denotes the largest integer not greater than $x$. More precisely,   
the first $N$ modes $\{\pm k_j\}$, $1\leqslant j \leqslant N$, are linearly unstable, since they give rise to exponentially growing and decaying waves of amplitudes~$O(\varepsilon e^{\pm \sigma_j t})$, where the growth rates~$\sigma_j$ are defined by
\begin{equation}
\label{eq:sigmas1}
\sigma_j=k_j\sqrt{4|a|^2-k^2_j}\,, \qquad
1\leqslant j \leqslant N,
\end{equation}
while the remaining modes are linearly stable, since they give rise to small oscillations of amplitude~$O(\varepsilon e^{\pm i \omega_j t})$, where 
\begin{equation}
\label{eq:sigmas2}
\omega_j=k_j\sqrt{k^2_j-4|a|^2}\,, \qquad
j>N.
\end{equation}

For the unstable part of the spectrum it is convenient to introduce the following notation:
\begin{equation}
\label{eq:phi1}
\phi_j=\arccos\left(\frac{k_j}{2|a|}\right)=\arccos\biggl(\frac{\pi}{L|a|}j\biggr),\qquad
0<\phi_j<\frac{\pi}{2}\,, \quad
1\leqslant j \leqslant N,
\end{equation}
implying that
\begin{equation}
\label{eq:phi1}
k_j=2|a|\cos\phi_j, \quad
\sigma_j=2|a|^2 \sin(2\phi_j), \qquad
1\leqslant j \leqslant N.
\end{equation}

In particular, if $\pi/|a| < L < 2\pi/|a|$, then $N=1$ and the solution is well approximated by a genus 2 exact solution on a Riemann surface with $O(\eps)$ handles, allowing one to describe, to leading order, the following exact recurrence of AWs in terms of elementary functions of the initial data.\\
Construct the following linear combinations of the Fourier coefficients of the unstable part of the initial perturbation:
\begin{equation}
\label{eq:alpha_beta}
\alpha =e^{-i\phi_1}\overline{c_1}-e^{i\phi_1}c_{-1},\quad \beta =e^{i\phi_1}\overline{c_{-1}}-e^{-i\phi_1}c_1,
\end{equation}
where, hereafter, $\bar f$ is the complex conjugate of $f$; then the solution of the Cauchy problem to leading order (up to $O(\epsilon)$ corrections), in the finite interval $0\le t \le T$,  reads as follows \cite{GS3}:
\beq\label{unif_sol_Cauchy_1}
\ba{l}
u(x,t)=\sum\limits_{m=0}^n {\cal A}\Big(x,t;\phi_1,x^{(m)},t^{(m)} \Big)
e^{i\rho^{(m)}} -a\frac{1-e^{4in\phi_1}}{1-e^{4i\phi_1}}e^{2i|a|^2t}, \ \ x\in [0,L], 
\ea
\eeq
where the parameters $x^{(m)},~ t^{(m)},~\rho^{(m)},~m\ge 0$, are defined in terms of the initial data by the following elementary functions 
\beq\label{parameters_1n}
\ba{l}
x^{(m)}=X^{(1)}+(m-1)\Delta X, \ \ t^{(m)}=T^{(1)} + (m-1)\Delta T, \\
X^{(1)}=\frac{\arg\alpha}{k_1} +\frac{L}{4}, \ \ \Delta X =\frac{\arg(\alpha\beta)}{k_1}, \ \ (\!\!\!\!\!\mod L),\\
T^{(1)}\equiv\frac{1}{\sigma_1}\log\left(\frac{2\sin^2(2\phi_1)}{\epsilon|\alpha|} \right)=
\frac{1}{\sigma_1}\log\left(\frac{\sigma^2_1}{2|a|^4\epsilon|\alpha|} \right), \\
\Delta T =\frac{1}{\sigma_1} \log\left(\frac{4\sin^4(2\phi_1)}{\epsilon^2 |\alpha\beta|}\right)=
\frac{1}{\sigma_1} \log\left(\frac{\sigma^4_1}{4|a|^8\epsilon^2 |\alpha\beta|}\right), \,  \\
\rho^{(m)}=2\phi_1+(m-1)4\phi_1 ,\\
n = \left\lceil \frac{T - T^{(1)}}{\Delta T} +\frac{1}{2} \right\rceil ,
\ea
\eeq
where $\lceil x \rceil$, $x\in\mathbb{R}$ denotes the smallest integer not smaller than $x$, and function~${\cal A}$ is the AB:
\begin{equation}
\label{eq:akh1}
\begin{gathered}
{\cal A}(x,t;\theta,X,T):= a~e^{2i|a|^2t}\frac{\cosh[\sigma(\theta)(t-T)+2i\theta]+
\sin\theta\cos[k(\theta)(x-X)]}{\cosh[\sigma(\theta)(t-T)]-
\sin\theta\cos[k(\theta)(x-X)]}\,,
\\[2mm]
k(\theta)=2|a|\cos\theta, \qquad
\sigma(\theta)=k(\theta)\sqrt{4|a|^2-k^2(\theta)}=2|a|^2\sin(2\theta),
\end{gathered}
\end{equation}
exact solution of NLS for all real parameters $\theta,X,T$ (see Figure 1).

The solution (\ref{unif_sol_Cauchy_1})-(\ref{eq:akh1}) shows an exact recurrence of AWs described, to leading order, by the Akhmediev breather, whose parameters change at each appearance according to (\ref{parameters_1n}). It is a very good example of a Fermi-Pasta-Ulam-Tsingou \cite{FPU} type recurrence without thermalisation (see \cite{GS2} for the analytic aspects of this recurrence in Fourier space). $X^{(1)}$ and $T^{(1)}$ are respectively the position and the time of the first appearance; $\Delta X$ is the $x$-shift of the position of the AW between two consecutive appearances, and $\Delta T$ is the recurrence time (the time between two consecutive appearances) (see Figure 1). Therefore $T^{(1)}$ and $\Delta T$ are the characteristic times of the AW recurrence.

\begin{figure}[H]
\centering
\includegraphics[width=2.5cm]{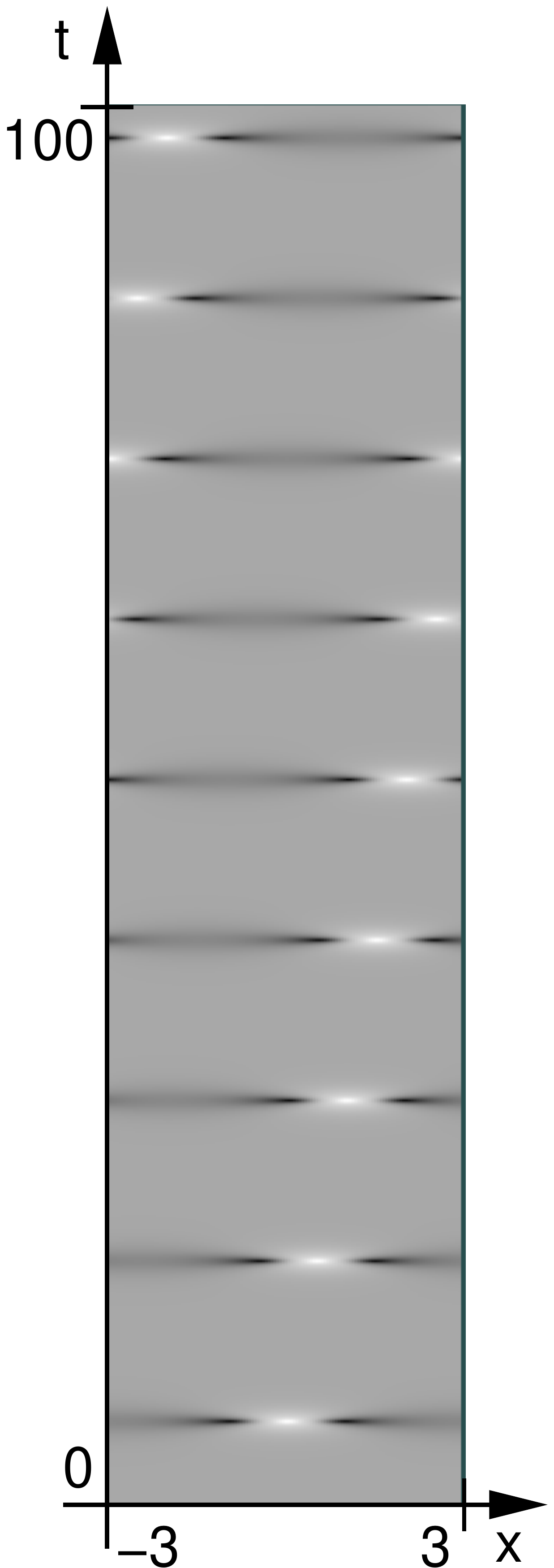}
\caption{The density plot of $|u(x,t)|$ with $-L/2\le x\le L/2$, $0\le t \le 100$, $L=6$, $\epsilon=10^{-4}$, $a=1$, with a generic initial condition $c_{-1}=0.3+0.3i$, $c_1=0.5$, obtained using the refined split-step method \cite{JR}, is in extremely good quantitative agreement with (\ref{unif_sol_Cauchy_1}),(\ref{parameters_1n}) \cite{GS1}. Hereafter brighter and darker colors in the figures correspond to higher and lower values of $|u(x,t)|$ respectively.}
\label{fig1}
\end{figure}
\vskip 5pt
We first remark that, although the initial condition is infinite dimensional, the AW recurrence is described, to leading and relevant order, by just the four real parameters $X^{(1)}$, $T^{(1)}$, $\Delta X$ and $\Delta T$. Four free real parameters appear, indeed, in the unstable part of the initial condition (\ref{eq:nls_cauchy1}) (the real and imaginary parts of $c_1$ and $c_{-1}$, or the real and imaginary parts of $\alpha$ and $\beta$).

Then we remark that this recurrence can be predicted from simple qualitative considerations (see \cite{GS1,GS2}). The unstable mode grows exponentially and becomes $O(1)$ at logarithmically large times, when one enters the nonlinear stage of MI, and one expects the generation of a transient, $O(1)$, coherent structure over the unstable background (the AW). Since the AB describes the one-mode nonlinear instability, it is the natural candidate to describe such a stage, at the leading order. Due again to MI, this AW is expected to be destroyed in a finite time interval, and one enters the third asymptotic stage, characterized, like the first one, by the background plus an $O(\eps)$ perturbation. This second linear stage is expected, due again to MI, to give rise to the formation of a second AW (the second nonlinear stage of MI), described again by the Akhmediev breather, but, in general, with different parameters. And this procedure iterates forever, in the integrable NLS model, giving rise to the generation of an infinite sequence of AWs described by different Akhmediev breathers. {\it Then the AW recurrence is a relevant effect of nonlinear MI in the periodic setting, and the finite gap method is the proper tool to give an analytic description of it}.

We also remark that formulas (\ref{eq:alpha_beta})-(\ref{eq:akh1}), in perfect quantitative agreement with the output of the corresponding numerical experiment \cite{GS1}, were successfully tested soon after their appearance in a nonlinear optics experiment \cite{PieranFZMAGSCDR}.

We end our remarks observing the the first attempt to apply the finite gap method to solve the NLS Cauchy problem on the segment, for periodic perturbations of the background, was made in \cite{Tracy}, but no connection was established between the initial data and the parameters of the $\theta$-function representation, and no description of the dynamics in terms of elementary functions was given.

Since dissipation can hardly be avoided in all natural phenomena involving AWs, a natural question arises at this point. 
What is the effect of a small dissipation on the NLS periodic AW dynamics?

The corresponding Cauchy problem of the AWs becomes
\beq\label{dissipative_NLS}
i u_t+u_{xx}+2|u|^2u=-i\nu u , \ \ 0<\nu\ll 1,
\eeq
$$
u(x,0)=a (1+\eps v(x)), \ \ a \in\CC, \ \ 0<\eps\ll 1,  \ \ v(x+L)=v(x) ,
$$
and its homogeneous background solution and linear growth rate are, respectively \cite{Segur}
\beq\label{background_dissipation}
\tilde u_0(x,t,\nu)=a \exp(-\nu t)\exp\left(i\frac{|a|^2}{\nu}\left(1-\exp(-2\nu t) \right) \right),
\eeq
and 
\beq\label{sigma_dissipation}
\tilde\sigma(t,\nu)=-\nu +k\sqrt{4 |a|^2 \exp(-2\nu t)-k^2}.
\eeq
It is known \cite{Segur} that, if the initial perturbation is sufficiently small, a small dissipation can quench the growth process before the nonlinear effects become relevant, stabilizing the MI. In any case, due to (\ref{sigma_dissipation}), instability is always canceled if the time interval of interest is sufficiently long \cite{Segur,KharifPeli3}. In this respect we remark that the presence of dissipation introduces another characteristic time (from (\ref{sigma_dissipation}))
\beq
T_{diss}=\frac{1}{\nu}\log\left(\frac{2 |a|}{k} \right)
\eeq
(the time at which the unstable mode $k$ becomes stable), 
and the stabilizing effect described in \cite{Segur} takes place when the initial perturbation is sufficiently small, and dissipation is strong enough to have $T_{diss}\le T^{(1)}$.  

But what happens in the interesting case in which dissipation is small and $T_{diss}\gg T^{(1)}$? And what happens if one is interested in experiments in which the time interval is not long enough to allow dissipation to cancel the instability? 

Partial answers to these questions came recently from the following water wave and numerical experiments. In \cite{Amin} two results were presented: i) an experiment in a tank showing that the highly non generic AB initial condition (\ref{eq:akh1}) evolves into a recurrence of ABs whose position is shifted by $\Delta X=L/2$ (half a period); ii) a numerical experiment showing that the same AB initial condition evolves, according to (\ref{dissipative_NLS}), into the same pattern, thus interpreting the result of the real experiment as the effect of dissipation. Soon after that, it was shown in \cite{Soto} that a real sinusoidal initial perturbation of the background, evolving numerically according to the focusing NLS equation perturbed by linear loss or gain terms, gives rise to a recurrence of ABs with shifts respectively $\Delta X=L/2$ or $\Delta X=0$. To the best of our knowledge, no theoretical quantitative explanation involving analytic formulas has been given so far to these real and numerical experiments.

In this paper we use some aspects of the exact theory presented in \cite{GS1,GS2}, in the simplest case of one unstable mode, together with few aspects of the theory of perturbations of soliton PDEs, to construct the proper analytic model describing quantitatively and in terms of elementary functions the effect of a small linear loss/gain on the dynamics of the NLS AWs arising from a generic periodic perturbation of the unstable background. In particular, we provide the theoretical explanation of all the above real and numerical experiments.

The paper is organized as follows. In \S 2 we present and discuss these analytic results, and \S 3 is devoted to their proof.

\section{Results}

In this Section we present the analytic results describing the $O(1)$ effects of a small linear loss/gain on the NLS AWs dynamics generated by an $O(\eps),~\eps\ll 1$ generic periodic perturbation of the unstable background, in the simplest possible case of one unstable mode.

The Cauchy problem of the AWs for the focusing NLS equation perturbed by a linear loss/gain term studied in this paper reads as follows:
\begin{equation}
\label{eq:nls2}
i u_t+u_{xx}+2|u|^2u=-i\nu u , \qquad \nu \in\RR, \ \  |\nu|\ll1, 
\end{equation}  
(if $\nu>0$ we have a small loss, if $\nu<0$ we have a small gain), $[0,T]$ denotes the time interval in which we construct the solution, 
\begin{equation}
\label{eq:nls_cauchy2}
u(x,0)=a(1+\varepsilon v(x)), \ \ a\in\CC, \ \ 
0<\varepsilon\ll 1, \ \ \ 
v(x+L)=v(x),
\end{equation}
where
\begin{equation}
\label{eq:nls_cauchy2b}
v(x)=\sum_{j=1}^{\infty}(c_j e^{i k_j x}+c_{-j}e^{-i k_j x}),\quad
k_j=\frac{2\pi}{L}j,\qquad \frac{\pi}{|a|}< L < \frac{2\pi}{|a|} \ \  (\Rightarrow \ \ N=1).
\end{equation}
We also assume that
\beq\label{a_const}
|\nu|T, \ |a|^2 |\nu|T^2\ll 1 ,
\eeq
and the meaning of these conditions can be explained observing that the background solution (\ref{background_dissipation}) of (\ref{eq:nls2}) behaves as follows
\beq
\ba{l}
\tilde u_0(t,\nu)=a\exp(-\nu t)\exp\left(2i|a|^2t\left(1-\nu t+O(\nu t)^2\right) \right).
\ea
\eeq
Therefore the amplitude and the oscillation frequency of the background slowly decrease if $\nu>0$ (loss), and slowly increase if $\nu<0$ (gain). The condition $|\nu| T\ll 1$ means that we can neglect the slow decay/growth of the amplitudes of the background and of the AWs; the condition $|a|^2 |\nu| T^2 \ll 1$ means that we can neglect the slow decay/growth of the oscillation frequency and its effects. In particular,  $a$ can be treated as a constant parameter under the above assumptions.

A more complete analytic study of the problem, in which also these effects are taken into account, together with the output of nonlinear optics experiments in which the complete theory is tested, will be presented in a subsequent paper.

It is important to remark at this point that the main reason for the $O(1)$ effects on the periodic AW dynamics due to a small loss or gain are consequence of the fact that the Cauchy problem (\ref{eq:nls2})-(\ref{eq:nls_cauchy2b}) involves two small parameters: $\nu$ and $\eps$. As we shall see below, {\it the proper comparison actually involves $\nu$ and $\eps^2$}.
\vskip 10pt

Under the above hypothesis, the AW recurrence described in (\ref{unif_sol_Cauchy_1})-(\ref{parameters_1n}) is significantly modified by the small loss/gain in the following way. The solution is still described by a recurrence of Akhmediev breathers 
\beq\label{unif_sol_Cauchy_2}
\ba{l}
u(x,t)=\sum\limits_{m=0}^{\tilde n} {\cal A}\Big(x,t;\phi_1,\tilde x^{(m)},\tilde t^{(m)} \Big)
e^{i\rho^{(m)}} -a\frac{1-e^{4in\phi_1}}{1-e^{4i\phi_1}}e^{2i|a|^2t}, \ \ x\in [0,L], 
\ea
\eeq
where $\tilde x^{(1)}= x^{(1)}$, $\tilde t^{(1)}=t^{(1)}$ are essentially the same as in (\ref{parameters_1n}), but now
\begin{equation}
  \label{eq:reps}
  \ba{l}
\Delta X_m:=\tilde x^{(m+1)} - \tilde x^{(m)} =\frac{\arg(Q_m)}{k_1}\ \ (\!\!\!\!\!\mod L), \\
\Delta T_m:=\tilde t^{(m+1)} - \tilde t^{(m)} =\frac{1}{\sigma_1} \log\left(\frac{4\sin^4(2\phi_1)}{\epsilon^2|Q_m|}\right)=\frac{1}{\sigma_1} \log\left(\frac{\sigma^4_1}{4|a|^8 \epsilon^2|Q_m|}\right),\, 
\ea
\end{equation}
with
\begin{equation}
\label{eq:reps2}
Q_m =\alpha\beta -\frac{\nu}{\epsilon^2}\frac{2\sin(2\phi_1)}{|a|^2} m=\alpha\beta -\frac{\nu}{\epsilon^2}\frac{\sigma_1}{|a|^4} m, \quad m\ge 1 ,
\end{equation}
$\alpha$ and $\beta$ are defined in (\ref{eq:alpha_beta}), and $\rho^{(m)}$ in (\ref{parameters_1n}). 

From the above elementary formulas (\ref{unif_sol_Cauchy_2})-(\ref{eq:reps2}), we distinguish the following cases (assuming that $|\alpha\beta|,\sigma_1,|a|=O(1)$).

\begin{itemize}

\item If $|\nu|\ll \epsilon^2$,  then $Q_m\sim \alpha\beta$ for every $m$, and there is no basic difference with the zero-gain/loss case presented in the introduction. In particular, if $\nu=0$, then $Q_m=\alpha\beta$ for every $m$, and formulas (\ref{unif_sol_Cauchy_2})-(\ref{eq:reps2}) coincide with formulas (\ref{unif_sol_Cauchy_1})-(\ref{parameters_1n}). 

\item If $|\nu |$ is approximately of order $\epsilon^2$, the AW first appearance is essentially not affected by loss/gain, but, after it, we have a transient, consisting of a few AW recurrences, in which $Q_m\to -(\nu/\eps^2 )(\sigma_1/|a|^4) m$ as $m$ increases, and the solution tends to one of the two asymptotic states characterized by the following elementary formulas (see the central picture of Figure 2): \\
1) \\
\begin{equation}
\label{eq:reps3}
\begin{gathered}
  \arg Q_m \rightarrow \pi \ \ \Rightarrow \ \ \Delta X_m\rightarrow L/2, \ \ \mbox{if} \ \nu>0 \ \ \mbox{(loss)},\\
  \arg Q_m \rightarrow 0 \ \ \Rightarrow \ \ \Delta X_m\rightarrow 0, \ \ \mbox{if} \ \nu<0 \ \ \mbox{(gain)}.
\end{gathered} 
\end{equation}
2) The recurrence time decreases as $m$ increases according to the elementary formula:
\begin{equation}
\label{eq:reps4}
\Delta T_m\to \frac{1}{\sigma_1} \log\left(\frac{2|a|^2\sin^3(2\phi)}{|\nu |m}\right)= \frac{1}{\sigma_1} \log\left(\frac{\sigma^3_1}{4|a|^4|\nu |m}\right). 
\end{equation}

\item If $|\nu |\gg \epsilon^2$, but the conditions $|\nu|T|,\nu|T^2 \ll 1$ are still fulfilled, then $Q_m\sim -(\nu/\eps^2 )(\sigma_1/|a|^4) m$, and, after the first AW appearance (essentially not affected by loss/gain) and without any transient, the solution enters immediately one of the above two recurrence patterns (see the right picture of Figure 2): 
\begin{equation}
\label{eq:SVA}
\begin{gathered}
  \arg Q_m= \pi \ \Rightarrow \ \Delta X_m= L/2, \ \mbox{if} \ \nu>0 \ \mbox{(loss)},\\
  \arg Q_m= 0 \ \Rightarrow \ \Delta X_m= 0, \ \mbox{if} \ \nu<0 \ \mbox{(gain)}, \\
  \Delta T_m= \frac{1}{\sigma_1} \log\left(\frac{\sigma^3_1}{4|a|^4 |\nu |m}\right).
\end{gathered} 
\end{equation}

\end{itemize}

These two asymptotic states describe {\it lower dimensional} recurrence patterns depending on just two real parameters defining their position in space-time, unlike the zero-loss/gain case, in which the recurrence depends on four real parameters (see the Introduction). In addition, while the difference between two consecutive recurrence times is finite:
\beq
|\Delta T_{m+1}-\Delta T_m |=\frac{1}{\sigma_1}\log\left(\frac{m+1}{m} \right),
\eeq
the relative difference $|\Delta T_{m+1}-\Delta T_m |/\Delta T_m$ is small, since $\nu$ is small. Therefore we have slowly varying lower dimensional asymptotic states that can be viewed as slowly varying attractors (SVAs), the {\bf loss/gain - SVAs}, completely ruled by the parameter $\nu$. Similar considerations are valid also in the case of a finite number of unstable modes, and will be presented in a subsequent paper.\\
\ \\
Also special initial conditions are described by formulas (\ref{unif_sol_Cauchy_2})-(\ref{eq:reps2}):

\begin{itemize}

\item If the initial condition is the highly non generic Akhmediev breather (\ref{eq:akh1}), as in the experiments in \cite{Amin}, then all the NLS spectral gaps are initially closed \cite{GS1,GS4}, $\beta=0$, and $Q_m=-(\nu/\eps^2 )(\sigma_1/|a|^4) m$ is real. It follows that we are basically as in the case $\nu\gg \epsilon^2$, and, after the first AW appearance, essentially not affected by loss/gain, the solution enters immediately the SVAs (\ref{eq:SVA}) (see Figure 3). Therefore an initial condition that theoretically should evolve, according to NLS, into the AB (i.e., with no recurrence), gives rise instead, in the presence of a small loss or gain, to an AW recurrence described by the above SVAs. The instability of the AB with respect to perturbations of NLS, due to small corrective terms or to the numerical scheme approximating NLS, has been already observed (see, for instance, \cite{AblowHerbst,AblowSchobHerbst,Amin,Soto,GS4}).

\item If the initial perturbation is real, then $\overline{c_{-1}}=c_1$ and $\alpha\beta=4\sin^2\phi_1 |c_1|^2>0$; consequently the AW dynamics for zero-loss/gain is characterized by $\Delta X=0$. If a small gain is present, $Q_m=\alpha\beta+(|\nu|/\eps^2 )(\sigma_1/|a|^4)m>0,~\forall~m$; it follows that $\Delta X_m=0,~\forall~m$, and $\Delta T_m$ decreases as $m$ increases. If, instead, a small loss is present, the real quantity $Q_m=\alpha\beta-(|\nu|/\eps^2 )(\sigma_1/|a|^4) m$ passes from positive values to negative values for a certain $\tilde m\ge 1$. Correspondingly, $\Delta X_m=0$ for $m<\tilde m$, and $\Delta X_m=L/2$ for $m\ge \tilde m$; in addition, as $m$ increases, $\Delta T_m$ increases if $m<\tilde m$, and decreases if $m\ge \tilde m$ (see the Figure 4, in which a detailed quantitative comparison between the theoretical formulas (\ref{unif_sol_Cauchy_2})-(\ref{eq:reps2}) and the output of a numerical experiment is made). 

\item If the initial perturbation is purely imaginary, then $\overline{c_{-1}}=-c_1$, and $\alpha\beta=-4\cos^2\phi_1 |c_1|^2<0$; consequently the AW dynamics for zero loss/gain is characterized by $\Delta X=L/2$. If a small loss is present, $Q_m=-(|\alpha\beta|+(|\nu|/\eps^2 )(\sigma_1/|a|^4)m)<0,~\forall~m$; it follows that $\Delta X_m=L/2,~\forall~m$, and $\Delta T_m$ decreases as $m$ increases. If, instead, a small gain is present, the real quantity $Q_m=-|\alpha\beta|+(|\nu|/\eps^2 )(\sigma_1/|a|^4)m$ passes from negative values to positive values for a certain $\tilde m\ge 1$. Correspondingly, $\Delta X_m=L/2$ for $m<\tilde m$, and $\Delta X_m=0$ for $m\ge \tilde m$; in addition, as $m$ increases, $\Delta T_m$ increases if $m<\tilde m$, and decreases if $m\ge \tilde m$.

\end{itemize}

We end this section with the following important remark.
\vskip 5pt
\noindent
{\it Since a small dissipation can hardly be avoided in all natural phenomena involving AWs, and since a very small dissipation induces $O(1)$ effects on the dynamics of periodic AWs, evolving into slowly varying lower dimensional asymptotic states (slowly varying loss/gain - attractors) analytically described in this paper through formulas (\ref{eq:SVA}),  we expect that these asymptotic states, together with their generalizations corresponding to more unstable modes, will have to play a basic role in the theory of AWs in nature.}    
\begin{figure}[H]
\centering
\includegraphics[width=3cm]{generic1.pdf} \ \includegraphics[width=3cm]{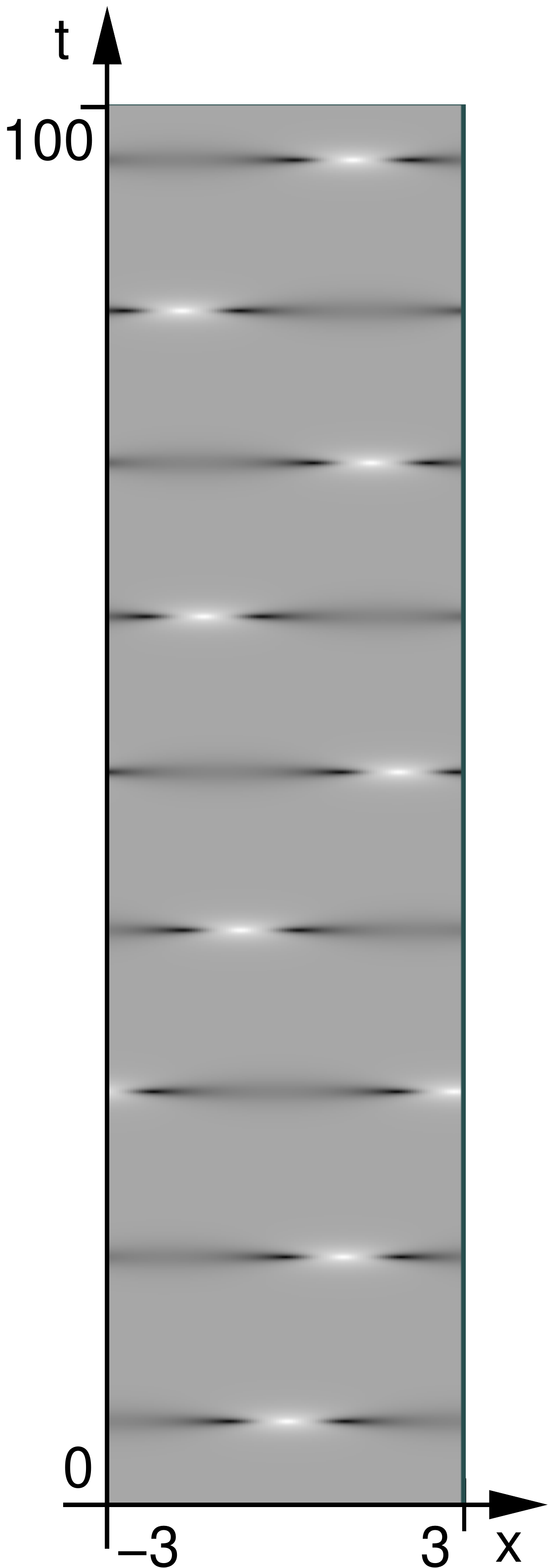} \ \includegraphics[width=3cm]{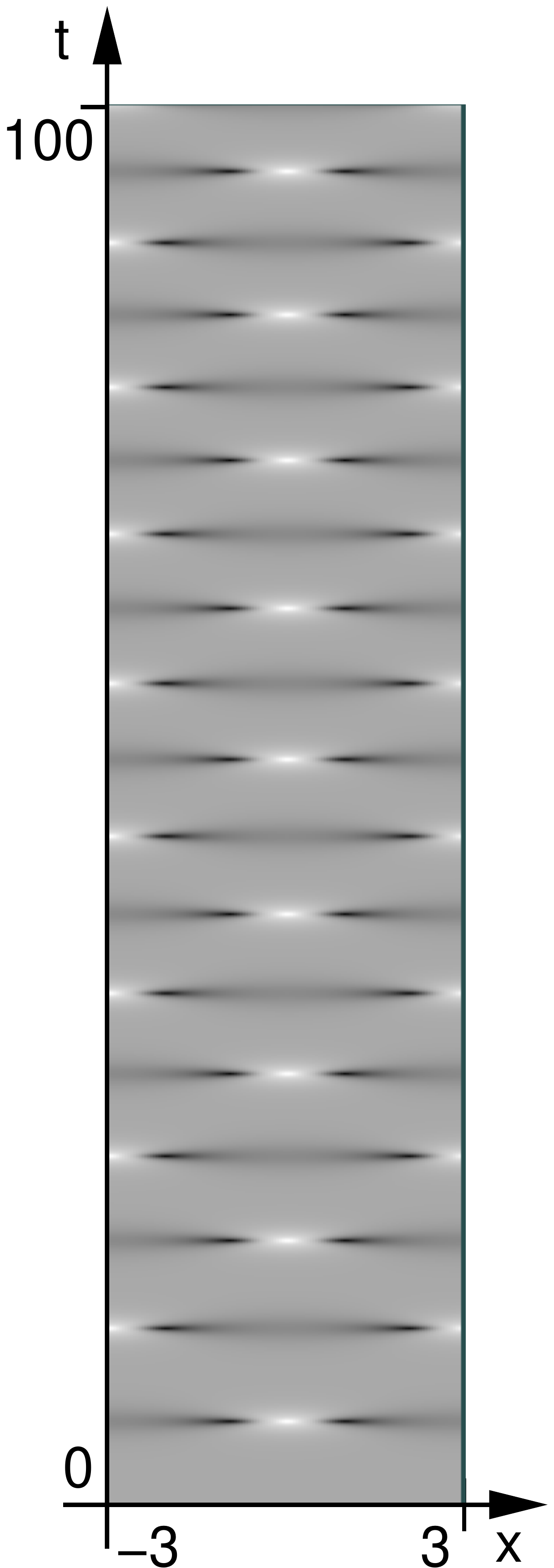}
\caption{The density plot of $|u(x,t)|$, with $-L/2\le x\le L/2$, $0\le t \le 100$, $L=6$, $\epsilon=10^{-4}$, $a=1$, for generic initial data: $c_1 = 0.5$ and $c_{-1}= 0.15-0.2 i$, obtained using the refined split-step method \cite{JR}. From left to right: $\nu=0$, $\nu=10^{-9}< \eps^2=10^{-8}$, and $\nu=10^{-5}\gg \eps^2$. In the left figure we have the usual AW recurrence described by formulas (\ref{unif_sol_Cauchy_1})-(\ref{parameters_1n}). In the central figure, the solution tends to the SVA (\ref{eq:SVA}) with $\Delta X_m\to L/2$,  after a relatively long transient. In the right figure, after the first appearance, the solution enters, without any transient, the SVA (\ref{eq:SVA}) with $\Delta X_m= L/2,~m\ge 1$. The first appearance is essentially the same in all the three cases.}
\label{fig2}
\end{figure}

\begin{figure}[H]
\centering
\includegraphics[width=3cm]{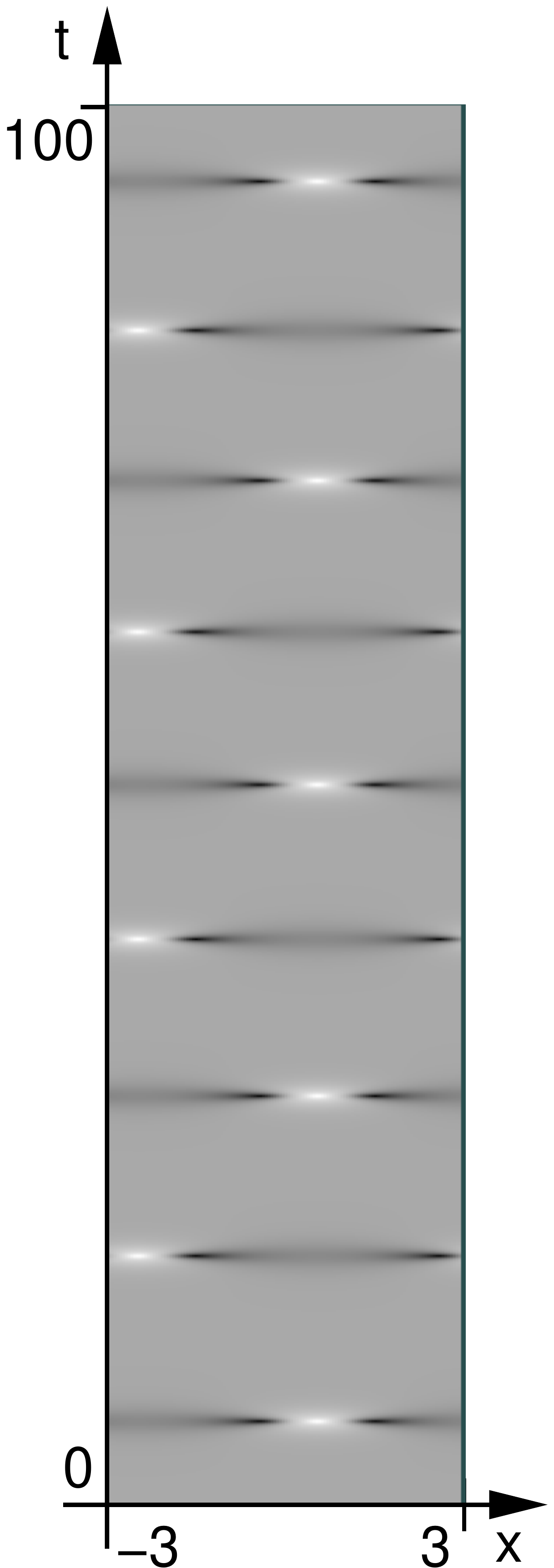} \ \includegraphics[width=3cm]{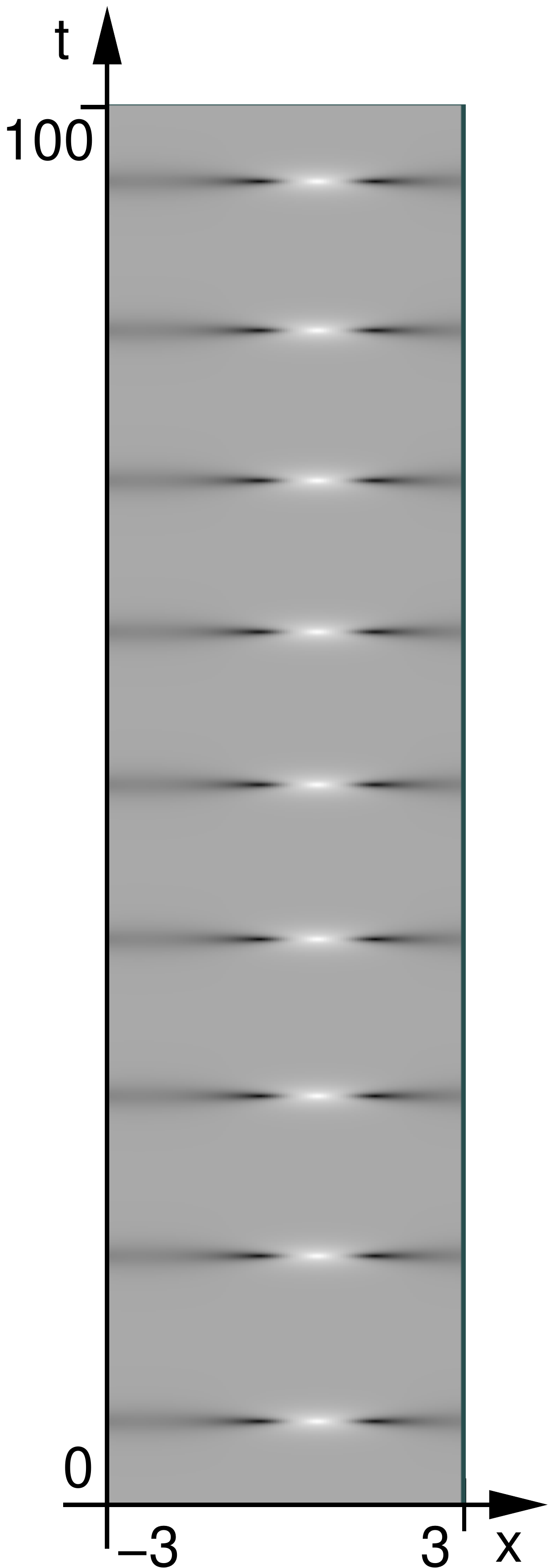} 
\caption{The density plot of $|u(x,t)|$ with $-L/2\le x\le L/2$, $0\le t \le 100$, $L=6$, $\epsilon=10^{-4}$, $a=1$, for the Akhmediev breather initial condition, corresponding to $c_1 = 0.223417921 + 0.113111515~i,~c_{-1} = -1.7601\cdot 10^{-10}+0.250419213~i$, obtained using the refined split-step method \cite{JR}. If $\nu=0$, the AW appears theoretically only once. If $\nu\ne 0$, after the first appearance, the solution enters, without any transient, the SVAs (\ref{eq:SVA}). $\nu=10^{-9}$ in the left figure, and $\nu=-10^{-9}$ in the right one. The first appearance is essentially the same in these two cases, and it essentially coincides with the AB appearance without loss/gain. }
\label{fig3}
\end{figure}
\begin{figure}[H]
\centering
\includegraphics[width=3cm]{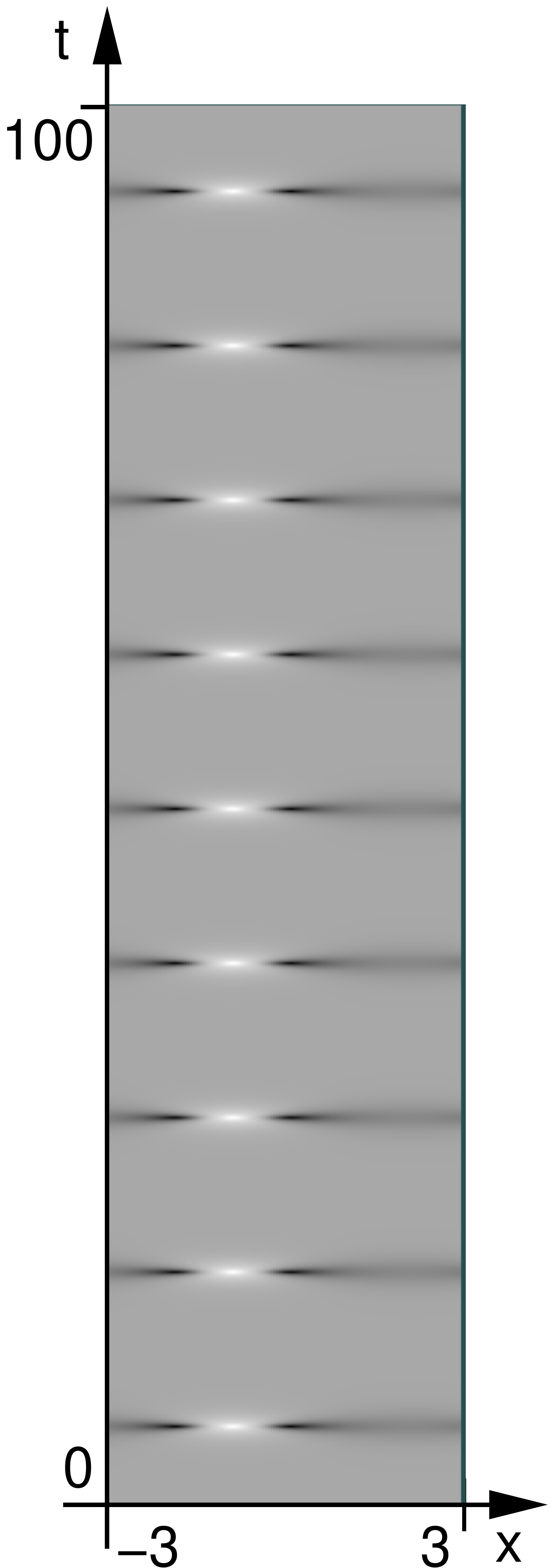} \ \includegraphics[width=3cm]{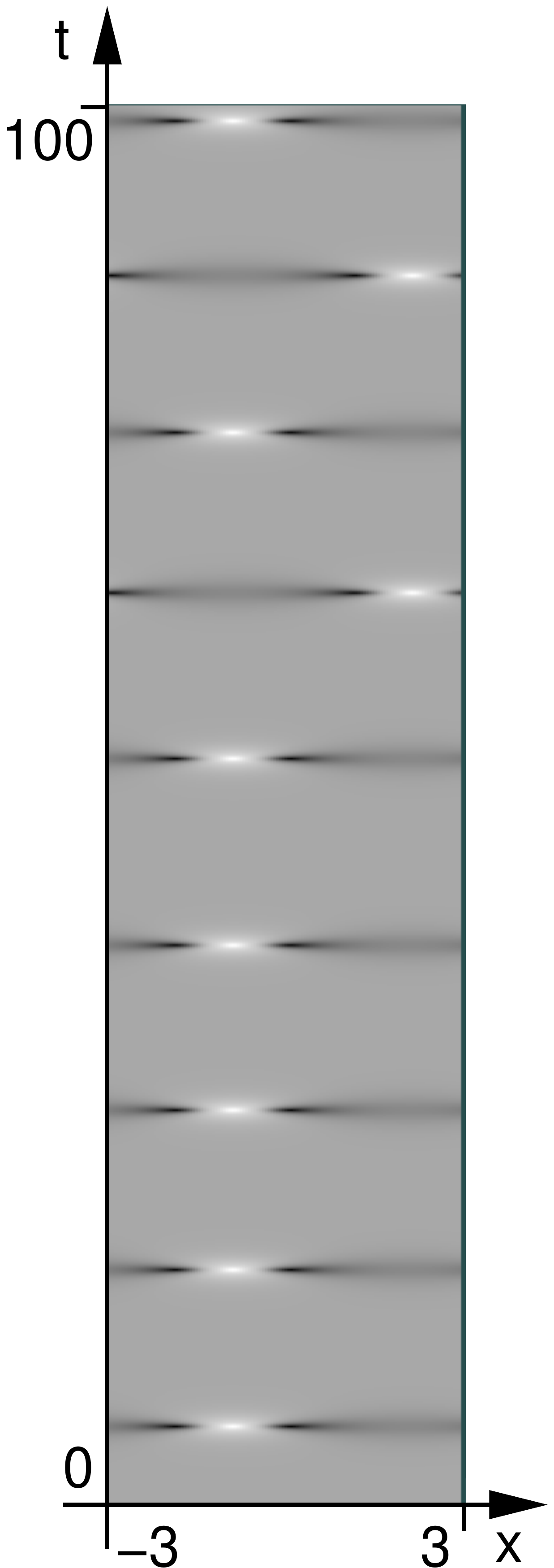} \ \includegraphics[width=3cm]{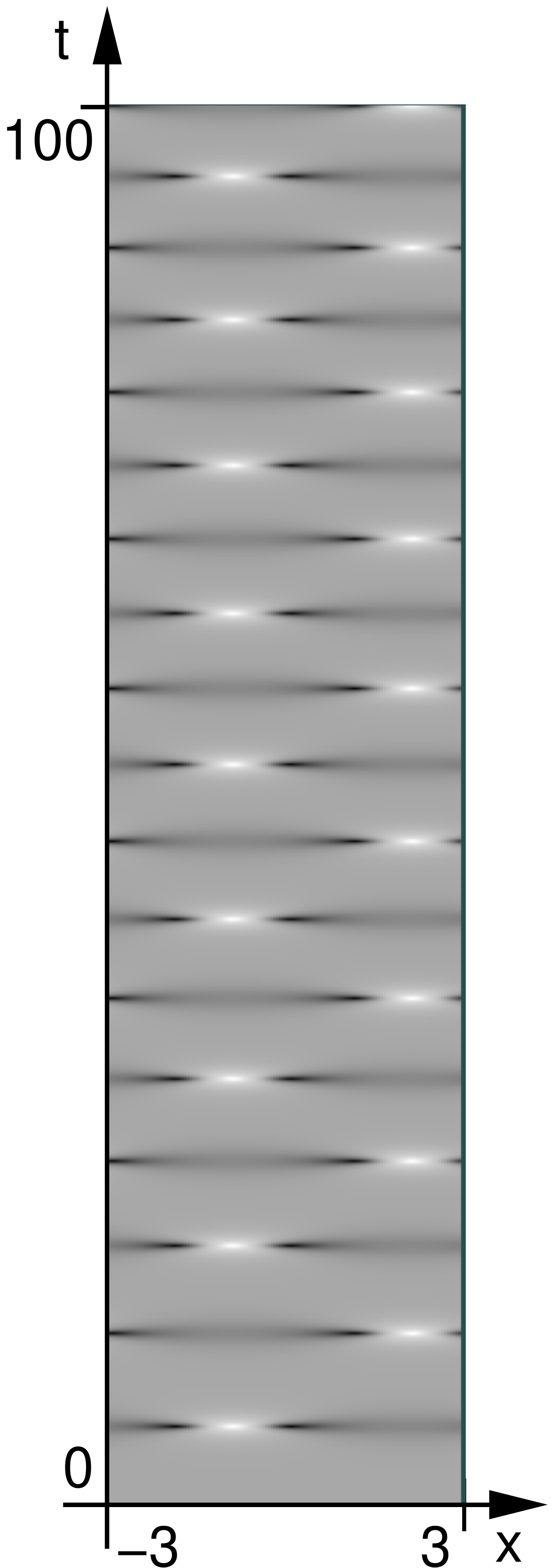} 
\caption{The density plot of $|u(x,t)|$ with $-L/2\le x\le L/2$, $0\le t \le 100$, $L=6$, $\epsilon=10^{-4}$, $a=1$,  for a real initial condition ($c_{-j}=\overline{c_j},~\forall j$), with $c_1=0.3+0.4 i$, obtained using the refined split-step method \cite{JR} with quadruple precision. Consequently $\alpha\beta>0$ and $Q_m$ is real.  Left picture: $\nu=0$, then $\Delta X=0$. Central picture: $\nu=10^{-9}$; then, for $\tilde m=6$, $Q_m$ changes its sign, from positive to negative values; correspondingly, $\Delta X_m$ switches from $0$ to $L/2$. Right picture: $\nu=10^{-5}\gg \eps^2$; then all $Q_m$ are negative and $\Delta X_m=L/2$ $\forall m$. The first appearance is essentially the same in all the three cases. For the central picture we also show the extremely good quantitative agreement between the theoretical predictions from formulas (\ref{unif_sol_Cauchy_2})-(\ref{eq:reps2}) and the numerical output:}
  \begin{center}
  \begin{tabular}{ll}
    $\tilde t^{(1)}=5.51209$ (theory) & $\tilde t^{(1)}=5.51208$ (numerics)\\
    $\Delta T_1=11.18230$ (theory) & $\Delta T_1=11.18230$ (numerics) \\
    $\Delta T_2=11.40337$ (theory) & $\Delta T_2=11.40338$ (numerics); \\
    $\Delta T_3=11.77375$ (theory) & $\Delta T_3=11.77376$ (numerics); \\
    $\Delta T_4=13.31847$ (theory) & $\Delta T_4=13.31848$ (numerics); \\
    $\Delta T_5=11.84989$ (theory) & $\Delta T_5=11.84988$ (numerics); \\
    $\Delta T_6=11.44140$ (theory) & $\Delta T_6=11.44142$ (numerics); \\
    $\Delta T_7=11.20765$ (theory) & $\Delta T_7=11.20766$ (numerics); \\
    $\Delta T_8=11.04319$ (theory) & $\Delta T_8=11.04320$ (numerics)
    \end{tabular}
\end{center}
\label{fig4}
\end{figure}

\section{Proof of the results}
In this section we prove the above results. To do it, we make use of the following ingredients. i) Some aspects of the deterministic theory of periodic AWs recently developed in \cite{GS1,GS2} using the finite-gap method; ii) few basic aspects of the theory of perturbations of soliton PDEs (developed in the infinite line case in \cite{Kaup1,Kaup2}), and in the finite-gap case in \cite{EFM}); iii) the classical Darboux transformations for NLS \cite{Yurov,Matveev0}.

The zero-curvature representation for the self-focusing NLS is:
\begin{equation}
\label{eq:lp-x}
\vec\Psi_x(\lambda,x,t)=\hat U(\lambda,x,t)\vec\Psi(\lambda,x,t),
\end{equation}
\begin{equation}
\label{eq:lp-t}
\vec\Psi_t(\lambda,x,t)=\hat V(\lambda,x,t)\vec\Psi(\lambda,x,t),
\end{equation}
\beq\label{def_U}
\hat U(\lambda,x,t)=\left [\begin {array}{cc} -i \lambda & i u(x,t)
\\\noalign{\medskip} i \overline{u(x,t)} & i \lambda\end {array}
\right ]=-i\lambda \sigma_3 +iU(x,t),
\eeq
$$
\sigma_3=\begin{bmatrix}  1 & 0 \\ 0 & -1 \end{bmatrix}, \ \ U = \begin{bmatrix}  0 & u(x,t) \\ \bar u(x,t) & 0  \end{bmatrix},
$$
$$
\hat V(\lambda,x,t)= \left[\begin {array}{cc} -2 i \lambda^2 + i u(x,t)\overline{u(x,t)} & 2 i \lambda u(x,t) - u_x(x,t)
\\\noalign{\medskip} 2 i \lambda \overline{u(x,t)} +\overline{u_x(x,t)} & 2 i \lambda^2- i u(x,t)\overline{u(x,t)} 
\end {array}
\right ],
$$
where 
$$
\vec\Psi(\lambda,x,t)= \left [\begin {array}{c} \Psi_1(\lambda,x,t) \\
\Psi_2(\lambda,x,t) \end {array}\right ].
$$
The linear problem (\ref{eq:lp-x}) can be rewritten as a spectral problem
\begin{equation}
\label{eq:lp-x2}
\LG\vec\Psi(\lambda,x,t)=\lambda \vec\Psi(\lambda,x,t),
\end{equation}
where 
$$
\LG=\left[ \begin {array}{cc} i\partial_x & u(x,t) \\ -  \overline{u(x,t)} & -i\partial_x \end {array}\right].
$$
It is essential that $\LG$ is not self-adjoint, and the spectrum of this problem typically contains complex points. 

\bigskip

Equivalently, one can write the following matrix analogue of the spectral problem (\ref{eq:lp-x})

\begin{equation}
\label{eq:lp-x3}
\hat\Psi_x= -i\sigma_3 \hat\Psi \Lambda + i U \hat\Psi, \ \ \ \ \ \Lambda = \begin{bmatrix}  \lambda & 0 \\ 0 & \lambda' \end{bmatrix},
\end{equation}
where the first and second columns of $\hat\Psi$ are vector solutions of (\ref{eq:lp-x}) for respectively the spectral parameters $\lambda$ and $\lambda'$.

\subsection{Finite-gap approximation of the AW Cauchy problem}

Here we summarize the aspects of the finite gap theory and of the results obtained in \cite{GS1,GS2} used in this work.  

If $\Psi(\lambda,x,t)$ is a fundamental matrix solution of (\ref{eq:lp-x}),(\ref{eq:lp-t}) such that $\Psi(\lambda,0,0)$ is the identity, then the monodromy matrix $T(\lambda)$ is the entire function of $\lambda$ defined by: $T(\lambda)=\Psi(\lambda,L,0)$. The eigenvalues and eigenvectors of $T(\lambda)$ are defined on a two-sheeted covering of the $\lambda$-plane. This Riemann surface $\Gamma$ is called the \textbf{spectral curve} and does not depend on time.  The eigenvectors of $T(\lambda)$ are the Bloch eigenfunctions
\begin{eqnarray}
\label{eq:bloch1}
\vec\Psi_x(\gamma,x,t) =U(\lambda(\gamma),x,t)\vec\Psi(\gamma,x,t),\hphantom{aaa\gamma \in \Gamma}\nonumber\\
\vec\Psi(\gamma,x+L,t)=e^{iLp(\gamma)} \vec\Psi(\gamma,x,t), \ \ \gamma \in \Gamma,
\end{eqnarray}
and $\lambda(\gamma)$ denotes the projection of the point $\gamma$ to the $\lambda$-plane.

The spectrum is exactly the projection of the set $\{\gamma\in\Gamma,\Im p(\gamma)=0 \}$ to the $\lambda$-plane. The end points of the spectrum are the branch points and the double points (obtained merging pairs of branch points) of $\Gamma$, at which $e^{iLp(\gamma)}=\pm 1$, or, equivalently, $\tr T(\lambda)=\pm 2$:
\begin{eqnarray}
\label{eq:branch2}
\vec\Psi(\gamma,x+L,t)=\pm \vec\Psi(\gamma,x,t), \ \  \gamma \in \Gamma.\nonumber
\end{eqnarray}

The spectral curve $\Gamma_0$ corresponding to the background (\ref{back1}) is rational, and a point $\gamma\in\Gamma_0$ is a pair of complex numbers $\gamma=(\lambda,\mu)$ satisfying the quadratic equation $\mu^2=\lambda^2+|a|^2$. The corresponding monodromy matrix:
\beq
\tr T_0(\lambda) = 2 \cos (\mu L)
\eeq
defines the branch points $(\lambda^{\pm}_0,\mu_0)=(\pm i|a|,0)$ and the resonant (double) points $(\lambda^{\pm}_n,\mu_n)=(\pm\sqrt{(n\pi/L)^2-|a|^2},n\pi/L)$, $n\in\ZZ,~n\ne 0$. Near the resonant points:
\begin{equation}
\label{eq:prop4.1}
\tr T_0(\lambda) = (-1)^n \left[2 -\frac{\lambda_n^2 L^4}{\pi^2 n^2}(\lambda-\lambda_n)^2+O((\lambda-\lambda_n)^4)    \right].
\end{equation}

A generic initial $O(\eps)$ perturbation (\ref{eq:nls_cauchy1}) of the background perturbs this spectral picture. The branch points $\lambda^{\pm}_0=\pm i|a|$ become $E_0= i|a|+O(\eps^2)$ and $\bar E_0= -i|a|+O(\eps^2)$, and all double points $\lambda^{\pm}_n$, $n\ge1$, generically split into a pair of square root branch points, generating infinitely many gaps. If $1\le n\le N$, where $N$ is the number of unstable modes, $\lambda^+_n$ splits into the pair of branch points ($E_{2n-1},E_{2n}$), and $\lambda^-_n$ into the pair of branch points ($\bar E_{2n-1},\bar E_{2n}$); if $n>N$, each $\lambda_n$ splits into a pair of complex conjugate eigenvalues. In the simplest case in which $N=1$, $E_1,E_2$ are the branch points obtained through the splitting of the excited mode $\lambda^+_1=i|a|\sin\phi_1=:\lambda_1$, and \cite{GS1}
\beq
  \label{eq:bps2}
  \ba{l}
  E_{l} =\lambda_1+(-1)^l\frac{\epsilon |a|^2}{2\lambda_1}\sqrt{\alpha\beta}+O(\epsilon^2) \\
=i|a|\left(\sin(\phi_1)+(-1)^{l+1}\frac{\eps}{2\sin\phi_1}\sqrt{\alpha\beta}+O(\epsilon^2)\right), \ \ l=1,2,
\ea
\eeq
so that 
\beq\label{def_gap}
(E_1-E_2)^2=-\frac{\eps^2 |a|^2 \alpha\beta}{\sin^2(\phi_1)}.
\eeq

From formulas (\ref{parameters_1n}) and (\ref{def_gap}), one can write the following relations, to leading order, between the unstable gap $E_1-E_2$ and the AW recurrence period $\Delta T$ and $x$-shift $\Delta X$:
\beq\label{gap_period}
\ba{l}
\Delta T= \frac{1}{\sigma_1}\log\left(\frac{\sigma_1^4}{4 |a|^6 \sin^2\phi_1  |E_1-E_2|^2} \right), \\
\Delta X=\frac{\arg(-(E_{1}-E_{2})^2)}{k_1}.
\ea
\eeq
In the perturbed case, the analogue of equation (\ref{eq:prop4.1}) is
\beq
\label{eq:taylor}
\tr T(\lambda) = \tr T(\tilde\lambda_1) + \frac{\tr T''(\tilde\lambda_1) }{2}(\lambda-\tilde\lambda_1)^2+O((\lambda-\tilde\lambda_1)^3),
\eeq
where $\tilde\lambda_1$ is the critical point for $\tr T(\lambda)$. It is easy to check that
$$
\tilde\lambda_1 -\lambda_1 =O\big(\max(\epsilon^2,\frac{(E_1-E_2)^2}{|a|^2},\nu T) \big),
$$
$$
\tr T(\tilde\lambda_1) -\tr T(\lambda_1) =O\big(\max(\epsilon^4,\frac{(E_1-E_2)^4}{|a|^4}, (\nu T)^2) \big).
$$
We also have 
$$
\tr T(\lambda)- \tr T_0 (\lambda) =O \big(\max(\epsilon^2,\frac{(E_1-E_2)^2}{|a|^2},\nu T) \big).
$$

Therefore, to leading order, one can replace in (\ref{eq:taylor}) $\tr T(\tilde\lambda_1)$ by $\tr T(\lambda_1)$, and $\tr T''(\tilde\lambda_1)$ by $\tr T''_0(\lambda_1)=\frac{2\lambda_1^2 L^4}{\pi^2}$, obtaining
\beq\label{taylor2}
\tr T(\lambda) \sim \tr T(\lambda_1) + \frac{\tr T''_0(\lambda_1) }{2}(\lambda-\lambda_1)^2=\tr T(\lambda_1)+\frac{\lambda_1^2 L^4}{\pi^2}(\lambda-\lambda_1)^2.
\eeq

In \cite{GS1,GS2} it was suggested to approximate the solution of the generic Cauchy problem, corresponding to infinitely many gaps (see Figure 5 (left picture)), by the finite-gap solution obtained closing all gaps near the real line, since they correspond to the stable modes and contribute to the solution to $O(\eps)$ (see Figure 5 (right picture)). If $N=1$, we obtain a genus 2 spectral curve with two $O(\eps)$ handles (see Figure 5 (right picture)).

\begin{figure}[H]
\centering
\includegraphics[width=6cm]{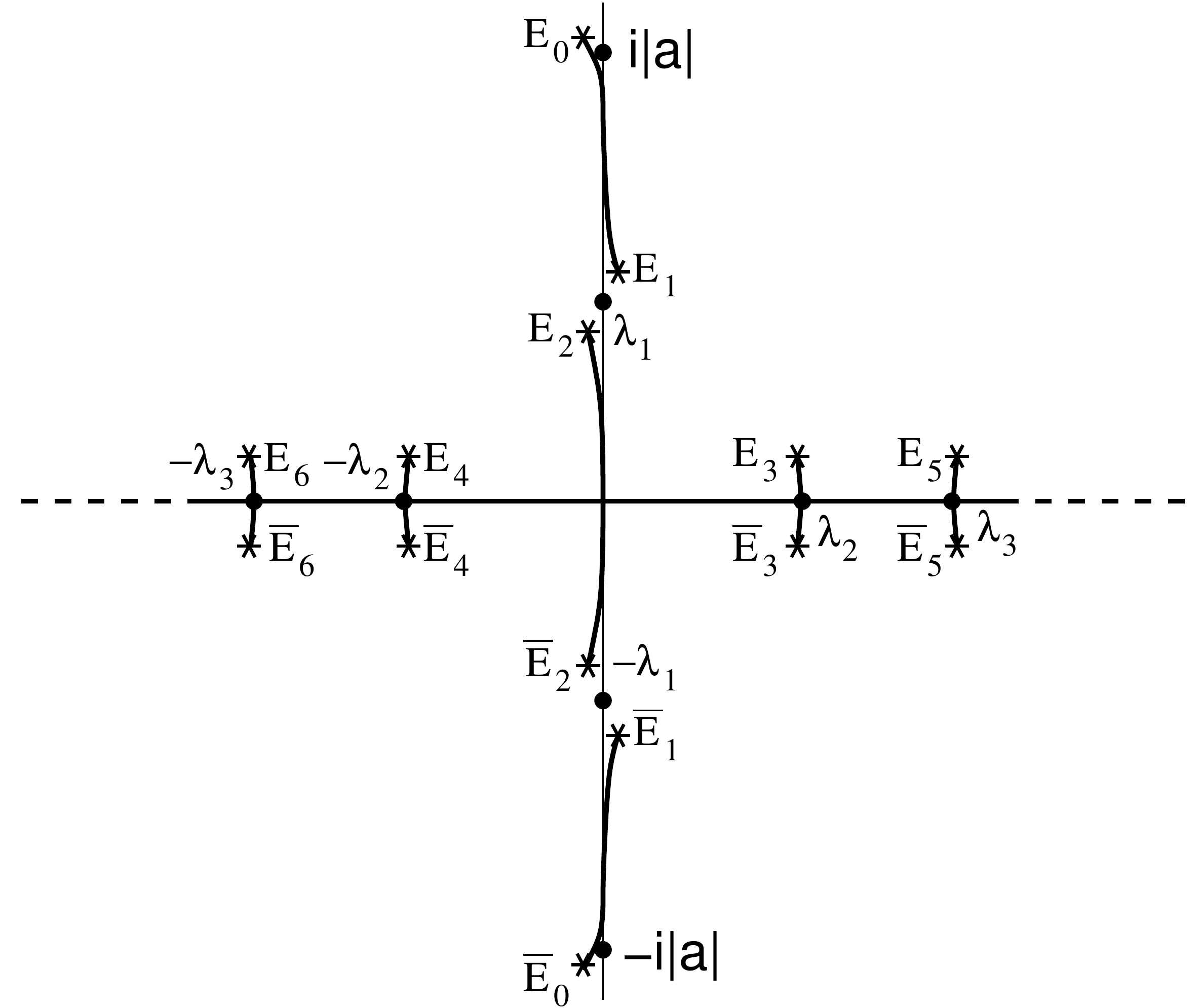}\includegraphics[width=6cm]{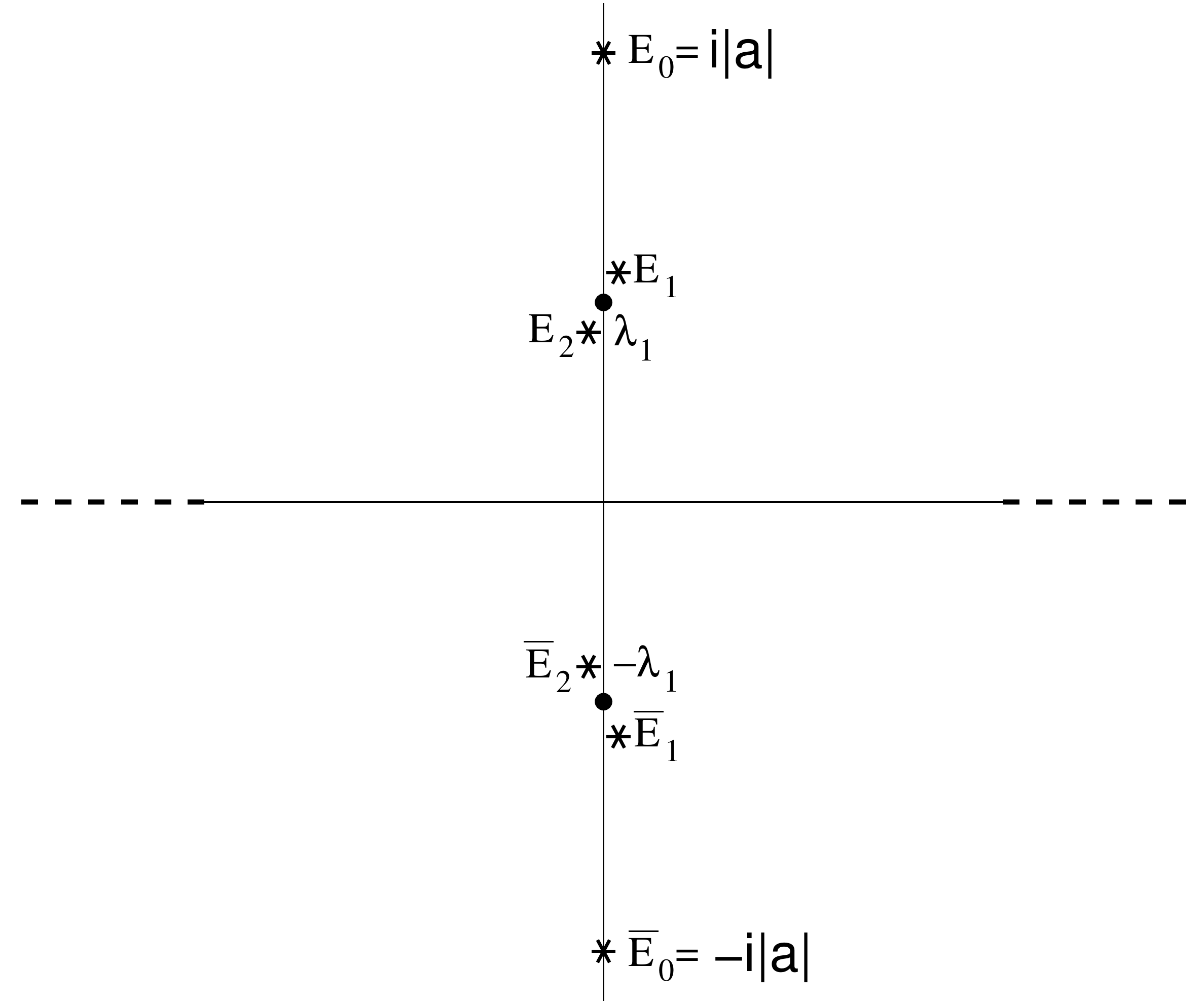}
\caption{On the left: the spectral curve generated by a small generic periodic perturbation of the background in case of one unstable mode.
On the right: the finite-gap genus 2  spectral curve approximating it. }
\label{fig5}
\end{figure}

\subsection{Variations}

If $u$ evolves according to NLS, the monodromy matrix and the spectral curve are constants of motion. Now we calculate how the monodromy matrix, and, as a corollary, the spectral curve evolve in time in the presence of a small loss or gain. The calculation of the variation of the monodromy matrix uses a standard formula from ordinary differential equations coming from the method of variation of constants.
A perturbation of the finite-gap potential $U^{(FG)}(x,t)$ in (\ref{eq:lp-x}), (\ref{def_U})
$$
U^{(FG)}(x,t) \rightarrow U^{(FG)}(x,t) + \delta U^{(FG)}(x,t)
$$
induces the following perturbation of the fundamental matrix solution of (\ref{eq:lp-x})
\beq\label{deltaPsi}
\ba{l}
\Psi^{(FG)}(\lambda,x,t) \rightarrow \Psi^{(FG)}(\lambda,x,t) +\delta \Psi^{(FG)}(\lambda,x,t),\\
\delta \Psi^{(FG)}(\lambda,x,t)=\Psi^{(FG)}(\lambda,x,t)\int\limits_0^x{\Psi^{(FG)}}^{-1}(\lambda,x',t)(i\delta U^{(FG)}(x',t))\Psi^{(FG)}(\lambda,x',t)dx'
\ea
\eeq
where we have assumed without loss of generality that $\delta \Psi^{(FG)}(\lambda,0,t)=0$.

Let ${\hat T}(\lambda,y,x,t)$ be the transition matrix of (\ref{eq:lp-x}) with respect to the variable $y$, normalized to be the identity matrix $E$ for $y=x$ \cite{Faddeev}:
\beq\label{def_G}
\ba{l}
{\hat T}_y(\lambda,y,x,t)=(-i\lambda \sigma_3+i U^{(FG)}(y,t)){\hat T}(\lambda,y,x,t), \\
{\hat T}(\lambda,x,x,t)=E.
\ea
\eeq
Then
\beq
{\hat T}(\lambda,y,x,t)=\Psi^{(FG)}(\lambda,y,t){\Psi^{(FG)}}^{-1}(\lambda,x,t),
\eeq
where $\Psi^{(FG)}(\lambda,x,t)$ is any fundamental matrix solution of  (\ref{eq:lp-x}), the following properties are satisfied
\beq\label{properties_G}
\ba{l}
{\hat T}(\lambda, y, x ,t) = {\hat T}^{-1}(\lambda, x, y ,t), \\
{\hat T}(\lambda, z, y ,t) {\hat T}(\lambda, y, x ,t) = {\hat T}(\lambda, z, x ,t), \\
{\hat T}(\lambda, x+L, L,t)={\hat T}(\lambda, x, 0,t), 
\ea
\eeq
and the monodromy matrix can be expressed in terms of the transition matrix as follows:
\beq\label{properties_G2}
T(\lambda,t)={\hat T}(\lambda,L,0,t)=\Psi^{(FG)}(\lambda,L,t){\Psi^{(FG)}}^{-1}(\lambda,0,t).
\eeq

From (\ref{deltaPsi}) and (\ref{properties_G2}) it follows that the induced variation of the monodromy matrix reads 
\beq\label{delta_trace}
\ba{l}
\delta{T}(\lambda,t)=i\Psi^{(FG)}(\lambda,L,t)\int\limits_0^L{\Psi^{(FG)}}^{-1}(\lambda,x,t)(\delta U^{(FG)}(x,t))\Psi^{(FG)}(\lambda,x,t)(\Psi^{(FG)})^{-1}(\lambda,0,t) dx\\
=i\int\limits_0^L {\hat T}^{(FG)}(\lambda,L,x,t)(\delta U^{(FG)}(x,t)){\hat T}^{(FG)}(\lambda,x,0,t)dx.
\ea
\eeq
Making use of the properties (\ref{properties_G}), we finally obtain
\beq\label{delta_equ}
\ba{l}
\delta \tr {T} (\lambda,t)= i \int\limits_0^L \tr \Big[{\hat T}(\lambda, x + L,L,t){\hat T} (\lambda,L, x,t) \delta U^{(FG)}(x,t){\hat T} (\lambda,x, 0,t) \times \\
\times ({\hat T})^{-1}(\lambda,x + L,L,t)  \Big] dx=i\int\limits_0^L \tr \left[{\hat T}(\lambda,x+ L,x,t)\delta U^{(FG)}(x,t)\right] d x.
\ea
\eeq
Choosing $\delta U=U_t dt=[i\sigma_3(U_{xx}+2U^3)-\nu U]dt$, then $\delta\tr T=(\tr T)_t dt$; substituting these variations into equation (\ref{delta_equ}) and taking into account that $\int\limits_0^L \tr \left[{\hat T}(\lambda,x+ L,x,t)\sigma_3\left(U^{(FG)}_{xx}(x,t)+2(U^{(FG)})^3(x,t)\right)\right] d x=0$, it follows that 
\beq
(\tr {T})_t(\lambda,t)=-i\nu\int\limits_0^L \tr \left[{\hat T}(\lambda,x+ L,x,t)U^{(FG)}(x,t)\right] d x.
\eeq
At last, integrating this equation over time, from $0$ to $t$, one obtains the variation of $\tr T$ in the time interval $[0,t]$:  
$$
\Delta \tr {T} (\lambda, t )= -i\nu \int\limits_0^t d\tilde t  \left[\int\limits_0^L \tr \left[{\hat T} (\lambda,L +x, x,\tilde t) U^{(FG)}(x,\tilde t) \right] d x \right]=
$$
$$
=-i \nu \int\limits_0^t d\tilde t  \left[ \int\limits_0^L  \bigg[{\hat T}_{21}(\lambda,x+L,x,\tilde t) u^{(FG)}(x,\tilde t) + {\hat T}_{12}(\lambda,x+L,x,\tilde t) \overline{u^{(FG)}(x, \tilde t)} \bigg]   d x \right].
$$

The calculation of the above integral with high genus theta-functions is very complicated. But, to leading order, this integral can be explicitly calculated in terms elementary functions using the following properties of this solution:
\begin{enumerate}
\item Near each AW appearance the solution is well approximated by the Akhmediev breather.
\item Far from the AW appearance, the integral over the $x$-period tends to zero exponentially in $t$. Therefore the integral over the finite time interval of each AW appearance can be well approximated by the integral over the whole line $-\infty< t < \infty$ of the Akhmediev solution.  
\end{enumerate}
We conclude that, to leading order,
$$
\Delta \tr T (\lambda, t )= n_{app} \nu J(\lambda), 
$$
where $n_{app}$ is the number of AW appearances in the time interval $[0,t]$, and
\beq\label{def_J}
J(\lambda)
=-i \int\limits_{-\infty}^{+\infty} dt  \left[ \int\limits_0^L  \bigg[{\hat T}_{21}(\lambda,x+L,x,t) u(x,t) + {\hat T}_{12}(\lambda,x+L,x,t) \overline{u(x,t)} \bigg] d x \right],
\eeq
where $u(x,t)$ is the Akhmediev breather and ${\hat T}(\lambda,x+L,x,t)$ is its transition matrix. Let us recall that, to calculate the variation of the curve we need both $\Delta \tr T (\lambda, t )$ and $J(\lambda)$ at the point $\lambda=\lambda_1$. To compute ${\hat T}(\lambda,x+L,x,t)$ it is convenient to  use the classical Darboux transformation for NLS.

\subsection{Darboux transformation of the constant background}

Darboux Transformations (DTs) are used to construct solutions of integrable PDEs from simpler solutions \cite{Matveev0}. Here we use the well-known DT of NLS to construct the Akhmediev breather and its eigenfunctions from the constant background solution. \\ 
\ \\
Proposition \cite{Yurov}. Let $\vec\Psi_0(\lambda,x,t)$ be a solution of (\ref{eq:lp-x}),(\ref{eq:lp-t}) for  $U=U_0=\begin{bmatrix}  0 & u_0(x,t) \\ \bar u_0(x,t) & 0  \end{bmatrix}$, and let $\hat\Phi_0$ be a fixed solution of (\ref{eq:lp-x3}) for  $U=U_0$, but with a fixed $\Lambda=\Lambda_1=\begin{bmatrix}  \lambda_{1} & 0 \\ 0 & \lambda'_{1} \end{bmatrix}$. Let $\tau$ be
\beq\label{def_tau}
\tau(x,t) = \hat\Phi_0(x,t) \Lambda_1 \hat\Phi_0^{-1}(x,t) ,
\eeq
implying that
$$
\tau_x = -i[\sigma_3,\tau]\tau + i[U_0,\tau].
$$
Then
\beq\label{dressed}
\vec\Psi(\lambda,x,t) = [\lambda  - \tau(x,t)]\vec\Psi_0(\lambda,x,t), \ \ U(x,t) = U_0(x,t) -[\sigma_3,\tau(x,t)]
\eeq
are the Darboux transformed  eigenfunction and potential satisfying (\ref{eq:lp-x}),(\ref{eq:lp-t}).

We apply this transformation on the constant background (\ref{back1}). As we have already seen, the unperturbed spectral curve $\Gamma_0$ is rational, and a point $\gamma\in\Gamma_0$ is a pair of complex numbers $\gamma=(\lambda,\mu)$ satisfying the equation $\mu^2=\lambda^2+|a|^2$. On the imaginary interval $-|a|\le \Im\lambda \le |a|$, $\Re\lambda=0$ we can write:
$$
\lambda = i|a| \sin(\phi), \ \ \mu = |a|\cos(\phi), \ \ \lambda+\mu = |a|e^{i\phi}.
$$

The Bloch eigenfunctions for the operator $\LG_0$ can be easily calculated explicitly:
\begin{equation}
\label{eq:bloch3}
\vec{\psi}^{\pm}(\gamma,x)=\left[\begin {array}{c}a e^{i|a|^2t} \\ \big[ \lambda(\gamma)\pm \mu(\gamma) \big] e^{-i|a|^2 t} \end {array} \right ] e^{\pm i\mu(\gamma) x \pm 2i \lambda(\gamma)\mu(\gamma)t },
\end{equation}
$$
{\LG_0} \psi^{\pm}(\gamma,x) = \lambda(\gamma)  \psi^{\pm}(\gamma,x),
$$
or, in a different normalization,
\begin{equation}
\label{eq:bloch4}
\vec{\psi}^{\pm}(\phi,x)=\left[\begin {array}{c}a e^{i|a|^2 t\mp i\phi/2} \\  \pm |a| e^{-i|a|^2 t \pm i \phi/2} \end {array} \right ] e^{\pm i|a|\cos(\phi)x \mp |a|^2\sin(2\phi)t}.
\end{equation}
Denote by $\vec q$ the special solution of (\ref{eq:lp-x}), for $u=u_0$, and $\lambda=\lambda_1$, obtained adding up the two vector solutions (\ref{eq:bloch4}):
\begin{equation}
\label{eq:q1}
\vec q(x,t) =
\left[
  \ba{l}
  q_1 \\
  q_2
  \ea
  \right]=
\left[\begin {array}{c}a e^{i|a|^2 t}\big(e^{-i\phi_1/2 + i|a|\cos(\phi_1)x -|a|^2 \sin(2\phi_1) t} + e^{i\phi_1/2 - i|a|\cos(\phi_1)x + |a|^2\sin(2\phi_1) t} \big)    \\ |a| e^{-i|a|^2 t}\big(e^{i\phi_1/2 + i|a|\cos(\phi_1)x -|a|^2 \sin(2\phi_1) t} - e^{-i\phi_1/2 - i|a|\cos(\phi_1)x +|a|^2 \sin(2\phi_1) t} \big) \end {array} \right ]
\end{equation}
\begin{equation}
\label{eq:q1}
= 2 \left[\begin {array}{c} ae^{i|a|^2t} \cos (\frac{k_1}{2} x-\phi_1/2 + i\frac{\sigma_1}{2} t) \\ i|a| e^{-i|a|^2t}  \sin (\frac{k_1}{2} x +\phi_1/2 + i\frac{\sigma_1}{2} t) 
\end{array}\right],
\end{equation}
where
$$
\lambda_{1}=i|a|\sin(\phi_1), \ \ \mu_1=|a|\cos(\phi_1), \ \ k_1 = 2|a|\cos(\phi_1), \ \ \sigma_1 = 2|a|^2 \sin(2\phi_1),
$$
and we assume $\phi_1$ to be real. The complex symmetry of (\ref{eq:lp-x}) implies that, since $\left[
  \ba{l}
  {q_1} \\
  {q_2}
  \ea
  \right]$ is solution of (\ref{eq:lp-x}) for $u=u_0$ and $\lambda=\lambda_1$, then $\left[
  \ba{c}
  -\overline{q_2} \\
  \overline {q_1}
  \ea
  \right]$ is solution of (\ref{eq:lp-x}) for $u=u_0$ and $\lambda=\overline{\lambda_1}=-i|a|\sin(\phi_1)$. Consequently
\begin{equation}
\label{eq:darb1}
\hat\Phi_0=\begin{bmatrix} q_1 & -\overline{q_2} \\ q_2 & \overline {q_1} \end{bmatrix}
\end{equation}
solves (\ref{eq:lp-x3}) for $U=U_0$ and $\Lambda_1= \begin{bmatrix}  \lambda_1 & 0 \\ 0 & \overline{\lambda_1} \end{bmatrix}$, 
and (from (\ref{def_tau}))
\begin{equation}
\label{eq:darb2}
\tau(x,t)=\frac{\lambda_1}{ \Den(x,t)}  \begin{bmatrix} q_1(x,t)\overline{q_1}(x,t)-  q_2(x,t)\overline{q_2}(x,t)  & 2 q_1(x,t)\overline{q_2}(x,t) \\ 2\overline{q_1}(x,t) q_2(x,t) & -q_1(x,t)\overline{q_1}(x,t)+q_2(x,t)\overline{q_2}(x,t)  \end{bmatrix},
\end{equation}
where
$$
\Den(x,t) = q_1(x,t)\overline{q_1(x,t)}+q_2(x,t)\overline{q_2(x,t)}=4|a|^2\left[\cosh(\sigma_1t)+\sin(\phi_1)\sin(k_1 x)\right].
$$
At last, the dressed potential reads
\beq\label{dressed_pot}
\ba{l}
u(x,t) =ae^{2i|a|^2t} -\frac{4\lambda_1}{Den(x,t)} q_1(x,t) \overline{q_2(x,t)}
=ae^{2i|a|^2t}\frac{\cosh[\sigma_1 t+2i\phi_1]-
\sin(\phi_1)\sin(k_1 x)}{\cosh(\sigma_1t)+
\sin(\phi_1)\sin(k_1 x)}.
\ea
\eeq
It coincides with the Akhmediev breather solution (\ref{eq:akh1}) after identifying $\phi_1=\theta$ and introducing suitable free parameters corresponding to the $x$ and $t$ translation symmetries of NLS.

If $u=u_0(x,t)$, then the corresponding transition matrix reads:
\begin{equation}
\label{eq:prop1}
{\hat T}_0(\lambda,y,x,t)=\begin{bmatrix} 
\cos(\mu (y-x)) -\frac{i\lambda}{\mu} \sin(\mu (y-x)) & \frac{ia}{\mu} \sin(\mu (y-x)) e^{2i|a|^2t} \\
\frac{i\bar a}{\mu} \sin(\mu (y-x)) e^{-2i|a|^2t} &  \cos(\mu (y-x)) +\frac{i\lambda}{\mu}  \sin(\mu (y-x))  
\end{bmatrix}=
\end{equation}
$$
=\frac{1}{\mu}\begin{bmatrix} 
|a|\cos(\mu (y-x)-\phi)  & ia \sin(\mu (y-x) ) e^{2i|a|^2t} \\
i\bar a \sin(\mu (y-x) ) e^{-2i|a|^2t} & |a| \cos(\mu (y-x)+\phi)  
\end{bmatrix} ,
$$
with
\begin{equation}
\label{eq:prop2}
\partial_{\lambda}{\hat T}_0(\lambda,y,x,t)=\frac{i\sin(\mu (y-x))}{\mu^3}  \begin{bmatrix} 
 -|a|^2     & -a \lambda  e^{2i|a|^2t} \\-\bar a \lambda e^{-2i|a|^2t} & |a|^2  
\end{bmatrix} +
\end{equation}
$$
+\frac{ \lambda (y-x) }{\mu^2}\begin{bmatrix} 
-|a|\sin (\mu (y-x)-\phi) & ia\cos(\mu (y-x) ) e^{2i|a|^2t} \\ i\bar a \cos(\mu (y-x) ) e^{-2i|a|^2t} & -|a|\sin (\mu (y-x)+ \phi) 
\end{bmatrix}.
$$
Consequently:
\begin{equation}
  \label{eq:prop3}
  \ba{l}
{\hat T}_0(\lambda_n,x+L,x,t)=(-1)^n E, \\
\partial_{\lambda}{\hat T}_0(\lambda,x+L,x,t)\bigg|_{\lambda=\lambda_n}=(-1)^n n \frac{ i\lambda_n \pi }{\mu_n^3}
\begin{bmatrix} 
 -\lambda_n &a e^{2i|a|^2t} \\ \bar a e^{-2i|a|^2t} & \lambda_n 
\end{bmatrix}.
\ea
\end{equation}

If $u$ is the potential (\ref{dressed_pot}), dressed from the background $u_0$, then the corresponding transition matrix reads (from (\ref{dressed}))
\beq\label{G_G0}
\ba{l}
{\hat T} (\lambda,y,x,t)=\Psi(\lambda,y,t)\Psi^{-1}(\lambda,x,t)=(\lambda-\tau(y,t))\Psi_0(\lambda,y,t)\Psi_0^{-1}(\lambda,x,t)(\lambda-\tau(x,t))^{-1}\\
=(\lambda-\tau(y,t)){\hat T}_0(\lambda,y,x,t)(\lambda-\tau(x,t))^{-1},
\ea
\eeq
where ${\hat T}_0$ is given in (\ref{eq:prop1}) and $\tau$ in (\ref{eq:darb2}). To evaluate (\ref{G_G0}) at $y=x+L$ and $\lambda=\lambda_1$, we use the following expansions for $\lambda$ near $\lambda_1$:
\beq
\ba{l}
{\hat T}_0(\lambda,x+L,x,t)={\hat T}_0(\lambda_1,x+L,x,t)+\frac{\partial {\hat T}_0(\lambda,x+L,x,t)}{\partial\lambda}|_{\lambda=\lambda_1}(\lambda-\lambda_1)+O(\lambda-\lambda_1)^2\\
=-E-\frac{ i\lambda_1 \pi }{\mu_1^3}
\begin{bmatrix} 
 -\lambda_1 & a e^{2i|a|^2t} \\ \bar a e^{-2i|a|^2t} & \lambda_1 
\end{bmatrix}(\lambda-\lambda_1)+O(\lambda-\lambda_1)^2, \\
\lambda-\tau(y,t)=\frac{\lambda_1-\overline{\lambda_1}}{Den(y,t)}\left[\ba{c}
  -\overline{q_2(y,t)}\\
  \overline{q_1(y,t)}
  \ea\right][-q_2(y,t),q_1(y,t)]+O(\lambda-\lambda_1), \\
(\lambda-\tau(x,t))^{-1}=\frac{1}{(\lambda-\lambda_1)Den(x,t)}\left[\ba{c}
  q_1(x,t)\\
  q_2(x,t)
  \ea\right][\overline{q_1(x,t)},\overline{q_2(x,t)}]+O(1),
\ea
\eeq
and the periodicity property $\hat\Phi_0(\lambda_1,x+L,t)=-\hat\Phi_0(\lambda_1,x,t)$. After some algebra one finally obtains
\begin{equation}\label{eq:prop5}
  \ba{l}
{\hat T} (\lambda_1,x+L,x,t) = - E +\frac{2 \pi i \lambda^2_1}{\mu_1^3}\frac{f(x,t)}{Den^2(x,t)}\begin{bmatrix} - \overline{q_2(x,t)} \overline{q_1(x,t)} &  - \overline{q_2(x,t)} \overline{q_2(x,t)}  \\ 
\overline{q_1(x,t)}  \overline{q_1(x,t)} &   \overline{q_1(x,t)}  \overline{q_2(x,t)}  \end{bmatrix}.\\
f(x,t)=a e^{2i|a|^2t}q^2_2(x,t)-\bar a e^{-2i|a|^2t}q^2_1(x,t)-2i|a|\sin\phi_1~ q_1(x,t) q_2(x,t).
\ea
\eeq
Using (\ref{eq:prop5}) we finally write explicitely the integral $J(\lambda_1)$, defined in (\ref{def_J}), describing the variation of $\tr T(\lambda_1)$ at each AW appearance:
\beq\label{J}
\ba{l}
J(\lambda_1)= -\frac{2\pi \sin^2\phi_1}{|a|\cos^3\phi_1}\int\limits_{-\infty}^{+\infty} dt\int\limits_0^Ldx \frac{f(x,t)g(x,t)}{Den^2(x,t)} ,\\
g(x,t)=u(x,t){\bar q_1(x,t)}^2-\bar u(x,t){\bar q_2(x,t)}^2.
\ea
\eeq

\subsection{The final formulas}

The following two important simplifications
\beq
\ba{l}
f(x,t)=a e^{2i|a|^2t}q^2_2(x,t)-\bar a e^{-2i|a|^2t}q^2_1(x,t)-2i|a|\sin\phi_1~ q_1(x,t) q_2(x,t)=-4 a|a|^2\cos^2(\phi_1),\\
g(x,t)=u(x,t){\bar q_1(x,t)}^2-\bar u(x,t){\bar q_2(x,t)}^2=4 \bar a |a|^2\left(\cos(2\phi_1)-\sin(\phi_1)\sin(k_1 x-i\sigma_1 t)\right),
\ea
\eeq
lead to the double integral
$$
J(\lambda_1)=2\pi\frac{|a|\sin^2\phi_1}{\cos\phi_1} \int\limits_{-\infty}^{+\infty} dt \int\limits_0^Ldx \frac{\cos(2\phi_1)-\sin(\phi_1)\sin(k_1 x-i\sigma_1 t)}{(\cosh(\sigma_1 t)+\sin(\phi_1)\sin(k_1 x))^2} =
$$
\beq\label{derivative_trace}
=2\pi^2\sin^2(\phi_1)\int\limits_{-\infty}^{+\infty}\frac{\cosh(\sigma_1 t)}{(\cosh^2(\sigma_1 t)-\sin^2(\phi_1))^{3/2}} dt 
\eeq
$$
=\frac{2\pi^2\sin^2(\phi_1)}{\sigma_1}\int\limits_{-\infty}^{\infty}\frac{d(\sinh(\sigma_1 t))}{(\sinh^2(\sigma_1 t)+\cos^2(\phi_1))^{3/2}} =\frac{\pi^2\sin\phi}{|a|^2\cos^3\phi}.
$$
The integration with respect to $x$ has been done using contour integration. The integration wrt $t$ is even more elementary. Therefore the variation $\Delta_1(\tr T (\lambda_1))$ of $\tr T(\lambda_1)$, due to a single appearance of the AW, is given by
\begin{equation}\label{variation}
  \ba{l}
  \Delta_1(\tr T (\lambda_1))=\nu J(\lambda_1)= \frac{\nu}{|a|^2} \frac{\pi^2\sin\phi_1}{\cos^3\phi_1}.
  \ea  
\end{equation}

Evaluating (\ref{taylor2}) at $\lambda=E_1$ and recalling that $E_1-E_2=2(E_1-\lambda_1)$, we have
\beq\label{trace}
\tr T (\lambda_1)\sim -2 + \frac{\sin^2(\phi_1)L^4}{4\pi^2}(E_1-E_2)^2\left(=
-2-\eps^2\frac{L^4}{4\pi^2}\alpha\beta, \ \mbox{at} \ t=0 \right).
\eeq
Then, due to (\ref{variation}) and (\ref{trace}), after each appearance:
\beq
\frac{\nu}{|a|^2} \frac{\pi^2\sin\phi_1}{\cos^3\phi_1}=\Delta_1(\tr T (\lambda_1))=\frac{L^2\sin^2\phi_1}{4\cos^2\phi_1}\Delta_1\left((E_1-E_2)^2\right).
\eeq
Therefore we have established that, after each appearance of the AW, the square of the gap varies of the same $O(\nu)$ quantity:
\beq
\Delta_1\left((E_1-E_2)^2\right)=4\nu\cot(\phi_1)
\eeq
with (see (\ref{def_gap}))
\beq
(E_1-E_2)^2\bigg|_{t=0}=-\frac{\eps^2 |a|^2 \alpha\beta}{\sin^2\phi_1}.
\eeq
We conclude that
\beq\label{gap_rotation}
(E^{(m)}_1-E^{(m)}_2)^2=-\frac{\eps^2 |a|^2 \alpha\beta}{\sin^2\phi_1}+4m\nu\cot\phi_1, \ \ m\ge 0,
\eeq
where
$E^{(m)}_1,E^{(m)}_2$ are the positions of the branch points of the gap after the $m^{th}$ AW appearance. This formula implies that, as $m$ increases and the AW dynamics tends to the SVAs (\ref{eq:SVA}), the gap tends to become horizontal if $\nu>0$ (loss), or vertical if  $\nu<0$ (gain) (see Figure 6). Let us remark that, in contrast with \cite{EFM}, the dynamics of the spectral curve is here essentially discrete. The spectral description of the two SVAs is therefore given by the elementary formula
\beq
(E^{(m)}_1-E^{(m)}_2)^2=4m\nu\cot\phi_1
\eeq
showing that the length of the gap grows through the law
\beq
|E^{(m)}_1-E^{(m)}_2 |=2\sqrt{m|\nu |\cot\phi_1}.
\eeq
\begin{figure}[H]
\centering
\includegraphics[width=6cm]{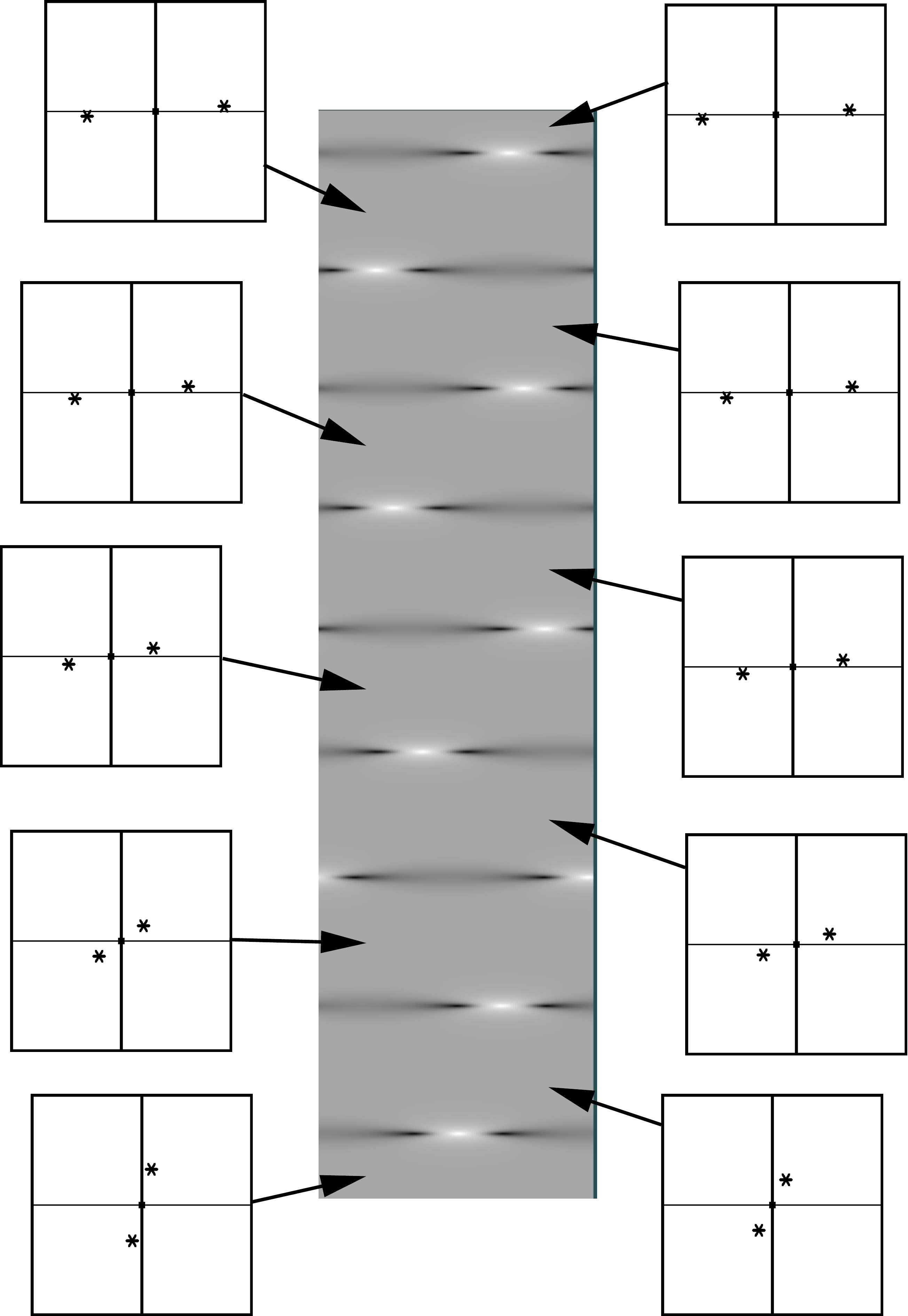}
\caption{The figure contains the numerical experiment illustrated in the central picture of Figure 2, together with the corresponding time evolution of the gap $E_1-E_2$, due to each AW appearance. It shows how the gap $E_1-E_2$ tends to become horizontal as the number of AW appearances increases, in the case of loss (in the case of gain, the gap would tend to become vertical). The quantitative agreement among the numerical output, the analytic formulas (\ref{unif_sol_Cauchy_2})-(\ref{eq:reps2}) describing the AW dynamics, and the analytic formula (\ref{gap_rotation}) describing the position of the gap after each AW appearance is extremely good.}
\label{fig6}
\end{figure}
It is also convenient to introduce the quantity $Q$ such that
\beq
\eps^2 \Delta_1 Q=-\sin^2 \! (\phi_1) ~\frac{\Delta_1 \! \left((E_1-E_2)^2\right)}{|a|^2}, \ \ \ 
Q\bigg|_{t=0}\! \! \! \! \! \! \! =\alpha\beta.
\eeq
Then also the variation of $Q$ after each appearance of the AW is constant:
\beq
\Delta_1 Q=Q_{m+1}-Q_m=-\frac{\nu}{|a|^2\eps^2}2\sin(2\phi_1)=-\frac{\nu\sigma_1}{|a|^4\eps^2}, \ \ \ \ \ Q_0=\alpha\beta
\eeq
implying (\ref{eq:reps2}).

\section{Acknowledgments} The work of P. G. Grinevich was supported by the Russian Science Foundation grant 18-11-00316. P. M. Santini was partially supported by the University La Sapienza, grant 2017.

We acknowledge useful discussions with F. Calogero, A. Chabchoub, C. Conti, E. DelRe, A. Degasperis, A. Gelash, A. Giansanti, D. Pierangeli, and, above all, with F. Briscese, with whom one of us (PMS) shared few years ago some calculations concerning the first linear stage of modulation instability and the first appearance of the AW in the presence of a small dissipation.


\begin{thebibliography}{99}

\bibitem{AblowHHShober} M.J. Ablowitz, J. Hammack, D. Henderson, C.M. Schober, ``Long-time dynamics of the modulational in stability of deep water waves'', \textit{Physica D,} \textbf{152--153} (2001), 416--433; doi:10.1016/S0167-2789(01)00183-X.

\bibitem{AblowHerbst} M.J. Ablowitz, B. Herbst, ``On homoclinic structure and numerically induced chaos for the nonlinear Schrodinger equation'', \textit{SIAM Journal on Applied Mathematics}, \textbf{50}:2 (1990), 339--351; doi:10.1137/0150021.

\bibitem{AL} M.J. Ablowitz, J.F. Ladik, ``Nonlinear differential-difference equations'', \textit{J. Math. Phys.,} \textbf{16}:3 (1975), 598--603; doi:10.1063/1.522558.

\bibitem{AM1} M.J. Ablowitz, Z.H. Musslimani, ``Integrable nonlocal nonlinear Schr\"odinger equation'', \textit{Phys. Rev. Lett.}, \textbf{110}:6 (2013), 064105; doi:10.1103/PhysRevLett.110.064105.

\bibitem{AblowSchobHerbst} M.J. Ablowitz, C.M. Schober, B.M. Herbst, ``Numerical Chaos, Roundoff Errors and Homoclinic Manifolds'', \textit{Phys. Rev. Lett.}, \textbf{71}:17 (1993), 2683--2686.10; doi:10.1103/PhysRevLett.71.2683.

\bibitem{AS} M.J. Ablowitz, H. Segur, \textit{Solitons and the Inverse Scattering Transform}, SIAM Studies in Applied Mathematics, Society for Industrial and Applied Mathematics, 1981, x+425 pp.

\bibitem{Akhmed0} N.N. Akhmediev, V.M. Eleonskii, and N.E. Kulagin, ``Generation of periodic trains of picosecond pulses in an optical fiber: exact solutions'', \textit{Sov. Phys. JETP}, \textbf{62}:5 (1985), 894--899.

\bibitem{Akhmed2} N.N. Akhmediev, V.M. Eleonskii, and N.E. Kulagin, ``Exact first order solutions of the Nonlinear Sch\"odinger equation'', \textit{Theor. Math. Phys}, \textbf{72}:2 (1987), 809--818.

\bibitem{Akhmed1} N.N. Akhmediev and V.I. Korneev, ``Modulation instability and periodic solutions of the nonlinear Schr\"dinger equation'', \textit{Theor. Math. Phys.}, \textbf{69}:2 (1986), 1089--1093.

\bibitem{Akhmed3} N.N. Akhmediev, ``Nonlinear physics: D\'ej\`a vu in optics'', \textit{Nature (London)}, \textbf{413} (2001), 267--268.

\bibitem{BDegaCW} F. Baronio, A. Degasperis, M. Conforti, S. Wabnitz, ``Solutions of the vector nonlinear Schrödinger equations: evidence for deterministic rogue waves'', \textit{Physical Review Letters} \textbf{109}:4 (2012), 44102; doi:10.1103/PhysRevLett.109.044102.

\bibitem{BBEIM} E.D. Belokolos, A.I. Bobenko, V.Z. Enolski, A.R. Its, V.B. Matveev, \textit{Algebro-geometric Approach in the Theory of Integrable Equations, Springer Series in Nonlinear Dynamics}, Springer, Berlin, 1994.

\bibitem{BF} T.B. Benjamin, J.E. Feir, ``The disintegration of wave trains on deep water. Part I. Theory'', \textit{Journal of Fluid Mechanics}, \textbf{27}:3 (1967) 417--430; doi:10.1017/S002211206700045X.

\bibitem{BT} M. Bertola and A. Tovbis, ``Universality for the focusing nonlinear Schr\"odinger equation at gradient catastrophe point: rational breathers and poles of the tritronque solution to Painlev\'e I'', \textit{Comm. Pure Appl. Math.}, \textbf{66}:5 (2013), 678--752; doi:10.1002/cpa.21445.

\bibitem{Talanov} V. I. Bespalov, V. I. Talanov, ``Filamentary structure of light beams in nonlinear liquids'', \textit{JETP Letters}, \textbf{3}:12 (1966), 307-310.

\bibitem{Biondini1} G. Biondini and G. Kovacic, ``Inverse scattering transform for the focusing nonlinear Schr\"odinger equation with nonzero boundary conditions'', \textit{J. Math. Phys.}, \textbf{55}:3, (2014), 031506; doi:10.1063/1.4868483.

\bibitem{Biondini2} G. Biondini, S. Li, D. Mantzavinos, ``Oscillation structure of localized perturbations in modulationally unstable media'', \textit{Phys. Rev. E}, \textbf{94}:6 (2016), 060201(R); doi:10.1103/PhysRevE.94.060201.

\bibitem{Bludov} Yu.V. Bludov, V.V. Konotop, N. Akhmediev, ``Matter rogue waves'', \textit{Physical Review A}, \textbf{80}:3 (2009), 033610; doi:10.1103/PhysRevA.80.033610.

\bibitem{Bortolozzo} U. Bortolozzo, A. Montina, F.T. Arecchi, J.P. Huignard, S. Residori, ``Spatiotemporal pulses in a liquid crystal optical oscillator'', \textit{Physical Review Letters}, \textbf{99}:2 (2007), 023901; doi:10.1103/PhysRevLett.99.023901.

\bibitem{CaliniEMcShober} A. Calini, N.M. Ercolani, D.W. McLaughlin, C.M. Schober, ``Mel'nikov analysis of numerically induced chaos in the nonlinear Schr\"odinger equation'', \textit{Physica D}, \textbf{89}(3--4) (1996), 227--260; doi:10.1016/0167-2789(95)00223-5.

\bibitem{CaliniShober1} A. Calini, C. Schober, ``Homoclinic chaos increases the likelihood of rogue wave formation'', \textit{Physics Letters A}, \textbf{298}(5--6) (2002), 335--349; doi:10.1016/S0375-9601(02)00576-5.

\bibitem{CaliniShober2} A. Calini, C.M. Schober, ``Dynamical criteria for rogue waves in nonlinear Schr\"odinger models'', \textit{Nonlinearity}, \textbf{25}:12 (2012) R99--R116; doi:10.1088/0951-7715/25/12/R99.

\bibitem{CHA_observP} A. Chabchoub, N. Hoffmann, N. Akhmediev, ``Rogue wave observation in a water wave tank'', \textit{Physical Review Letters}, \textbf{106}:20 (2011), 204502; doi:10.1103/PhysRevLett.106.204502.

\bibitem{Coppini} F. Coppini, P.M. Santini, ``
 Modulation instability for the Ablowitz-Ladik equations: exact solutions, periodic Cauchy problem, and rogue wave recurrence. I'', Preprint 2019 (in preparation).

\bibitem{DegaLomb0} A. Degasperis, S. Lombardo, ``Rational solitons of wave resonant interaction models'', Phys. Rev. E \textbf{88}, 052914 (2013).

\bibitem{DegaLomb} A. Degasperis, S. Lombardo, ``Integrability in action: solitons, instability and rogue waves, in Rogue and Shock Waves in non linear Dispersive Waves'', \textit{Lecture Notes in Physics}, M. Onorato, S. Resitori, F. Baronio (Eds.), 2016; http://www.springer.com/us/book/9783319392127.

\bibitem{DegaLombSommacal} A. Degasperis, S. Lombardo, M. Sommacal, ``Integrability and linear stability of nonlinear waves'', \textit{Journal of Nonlinear Science}, \textbf{28}:4 (2018), 1251--1291; doi:10.1007/s00332-018-9450-5.

\bibitem{DegaLombSommacal2} A. Degasperis, S. Lombardo, M. Sommacal, ``Rogue Wave Type Solutions and Spectra of Coupled Nonlinear Schrödinger Equations'', Fluids 2019, 4, 57. doi:10.3390/fluids401005. Open access: https://www.mdpi.com/2311-5521/4/1/57/pdf 

\bibitem{Dubrovin} B.A. Dubrovin, ``Inverse problem for periodic finite-zoned potentials in the theory of scattering'', \textit{Funct. Anal. Appl.}, \textbf{9}:1 (1975), 61--62; doi:10.1007/BF01078183.

\bibitem{Dysthe} K.B. Dysthe, K. Trulsen, ``Note on Breather Type Solutions of the NLS as Models for Freak-Waves'', \textit{Physica Scripta}, \textbf{T82}, (1999) 48--52; doi:10.1238/Physica.Topical.082a00048.

\bibitem{EKT} G.A. El, E.G. Khamis, A. Tovbis, ``Dam break problem for the focusing nonlinear Schr\"odinger equation and the generation of rogue waves'', \textit{Nonlinearity}, \textbf{29}:9 (2016), 2798--2836; doi:10.1088/0951-7715/29/9/2798.

\bibitem{EFM} N. Ercolani, M.G. Forest, D.W. McLaughlin, (1984) ``Modulational Stability of Two-Phase Sine-Gordon Wavetrains'', \textit{Studies in Applied Mathematics}, \textbf{71}(2), 97--101 (1984); doi:10.1002/sapm198471291. 

\bibitem{Ercolani} N. Ercolani, M.G. Forest, D.W. McLaughlin, ``Geometry of the modulation instability Part III: homoclinic orbits for the periodic Sine-Gordon equation'', \textit{Physica D}, \textbf{43}(2-3) (1990), 349--384; doi:10.1016/0167-2789(90)90142-C.

\bibitem{Faddeev}  L.D. Faddeev, L.A. Takhtajan, \textit{Hamiltonian methods in the theory of solitons}, Classics in Mathematics, Springer, Berlin, 2007, x+592 pp.

\bibitem{FL} M.G. Forest, Jong-Eao Lee, ``Geometry and Modulation Theory for the Periodic Nonlinear Schrodinger Equation'', in book: \textit{Oscillation Theory, Computation, and Methods of Compensated Compactness, The IMA Volumes in Mathematics and Its Applications}, vol 2. Springer, New York, NY (1986), 35--69.

\bibitem{FPU} G. Gallavotti (Ed.), ``The Fermi-Pasta-Ulam Problem: A Status Report'', \textit{Lecture Notes in Physics}, Vol. 728, Springer, Berlin Heidelberg, 2008; doi:10.1007/978-3-540-72995-2.

\bibitem{Gel} A.A. Gelash, ``Formation of rogue waves from a locally perturbed condensate'', \textit{Physical Review E}, \textbf{97} (2018), p. 022208; doi:10.1103/PhysRevE.97.022208.

\bibitem{GT} R.H.J. Grimshaw, A. Tovbis, ``Rogue waves: analytical predictions'', \textit{Proc. Roy. Soc. A}, \textbf{469}(2157) (2013), 20130094; doi:10.1098/rspa.2013.0094.

\bibitem{GS1} P.G. Grinevich, P.M. Santini, ``The finite gap method and the analytic description of the exact rogue wave recurrence in the periodic NLS Cauchy problem. 1'', \textit{Nonlinearity}, \textbf{31}:11 (2018), 5258--5308; doi:10.1088/1361-6544/aaddcf.

\bibitem{GS2} P.G. Grinevich, P.M. Santini: ``The finite-gap method and the periodic NLS Cauchy problem of anomalous waves for a finite number of unstable modes'', Russian Math. Surveys 74:2 211-263 (2019). DOI: https://doi.org/10.1070/RM9863.

\bibitem{GS3} P.G. Grinevich, P.M. Santini, ``The exact rogue wave recurrence in the NLS periodic setting via matched asymptotic expansions, for 1 and 2 unstable modes'', \textit{Physics Letters A}, \textbf{382}:14 (2018), 973--979; doi:10.1016/j.physleta.2018.02.014.

\bibitem{GS4} P.G. Grinevich, P.M. Santini: ``Numerical instability of the Akhmediev breather and a finite gap model of it'', arXiv:1708.00762. To appear in: V. M. Buchstaber et al. (eds.), Recent developments in Integrable Systems and related topics of Mathematical Physics, PROMS, Springer (2019).

\bibitem{GS5} P.G. Grinevich, P.M. Santini: ``Phase resonances of the NLS rogue wave recurrence in the quasi-symmetric case'', \textit{Theoretical and Mathematical Physics}, \textbf{196}:3 (2018), 1294--1306; doi:10.1134/S0040577918090040.

\bibitem{HendersonPeregrine} K.L. Henderson, D.H. Peregrine, J.W. Dold, ``Unsteady water wave modulations: fully nonlinear solutions and comparison with the nonlinear Schr\"odinger equtation'', \textit{Wave Motion}, \textbf{29}:4 (1999), 341--361; doi:10.1016/S0165-2125(98)00045-6.

\bibitem{Hirota0} R. Hirota, ``Exact Solution of the Korteweg-de Vries Equation for Multiple Collisions of Solitons'', \textit{Phys. Rev. Lett.}, \textbf{27}:18 (1971), 1192-1994; doi:10.1103/PhysRevLett.27.1192.

\bibitem{Hirota} R. Hirota, ``Direct Methods for Finding Exact Solutions of Nonlinear Evolution Equations'', in book: \textit{Solitons. ed. Robin K. Bullough, Philip J. Caudrey}, \textit{Lecture Notes in Mathematics}, Vol. 515, Springer, New York, 1976, 157--176; doi:10.1007/BFb0081162.

\bibitem{Kaup1} D.J.Kaup, A perturbation expansion for the Zakharov-Shabat inverse scattering transform, SIAM J. Appl. Math. {\bf 31} 121 (1976).
D.J.Kaup, Closure of the squared Zakharov-Shabat eigenstates, J. Math. Anal. Appl. {\bf 54}, 849–64 (1976).

\bibitem{Kaup2} D.J. Kaup, Integrable systems and squared eigenfunctions, Theor. Math. Phys. {\bf 159} 806-18 (2009) (Proc. the Workshop ``Nonlinear Physics: Theory and Experiment: V'')

\bibitem{KI} T. Kawata, H. Inoue, ``Inverse scattering method for the nonlinear evolution equations under nonvanishing conditions'', \textit{J. Phys. Soc. Japan}, \textbf{44}:5 (1978), 1722--1729; doi:10.1143/JPSJ.44.1722.

\bibitem{ItsMatveev} A.R. Its, V.B. Matveev, ``Hill's operator with finitely many gaps'', \textit{Funct. Anal. Appl.}, \textbf{9}:1 (1975), 65--66; doi:10.1007/BF01078185.

\bibitem{ItsKotlj} A.R. Its, V.P. Kotljarov, ``Explicit formulas for solutions of a nonlinear Schr\"odinger equation'', \textit{Dokl. Akad. Nauk Ukrain. SSR Ser. A}, \textbf{1051}, 965--968 (1976).

\bibitem{ItsRybinSall} A.R. Its, A.V. Rybin, M.A. Sall, ``Exact integration of nonlinear Schr\"odinger equation'', \textit{Theor. Math. Phys.}, \textbf{74} (1988), 20--32; doi:10.1007/BF01018207.

\bibitem{JR} J. Javanainen, J. Ruostekoski, ``Symbolic calculation in development of algorithms: split-step methods for the Gross-Pitaevskii equation'', J. Phys. A  \textbf{39}:12 (2006), L179-L184; ``Split-step Fourier methods for the Gross-Pitaevskii equation'', 2004, 3 pp.,  ArXiv:cond-math/0411154.

\bibitem{KharifPeli1} C. Kharif, E. Pelinovsky, ``Physical mechanisms of the rogue wave phenomenon'', \textit{Eur. J. Mech. B/ Fluids J. Mech.}, \textbf{22}:6 (2004), 603--634; doi:10.1016/j.euromechflu.2003.09.002.

\bibitem{KharifPeli2} C. Kharif, E. Pelinovsky, ``Focusing of nonlinear wave groups in deep water'' \textit{JETP Lett.}, \textbf{73} (2011), 170--175.

\bibitem{KharifPeli3}  C. Kharif, E. Pelinovsky, and A. Slunyaev, {\it Rogue Waves in the Ocean}, Springer-Verlag Berlin Heidelberg 2009.


\bibitem{KFFMDGA_observP} B. Kibler, J. Fatome, C. Finot, G. Millot, F. Dias, G. Genty, N. Akhmediev, J. Dudley, ``The Peregrine soliton in nonlinear fibre optics'', \textit{Nature Physics}, \textbf{6}:10 (2010), 790--795; doi:10.1038/nphys1740.

\bibitem{Amin} O. Kimmoun, H.C. Hsu, H. Branger, M.S. Li, Y.Y. Chen, C. Kharif, M. Onorato, E.J.R. Kelleher, B. Kibler, N. Akhmediev, A. Chabchoub, ``Modulation Instability and Phase-Shifted Fermi-Pasta-Ulam Recurrence'', \textit{Scientific Reports}, \textbf{6}, Article number: 28516 (2016), doi:10.1038/srep28516.

\bibitem{Krichever} I.M. Krichever, ``Methods of algebraic Geometry in the theory on nonlinear equations'', \textit{Russian Math. Surv.}, \textbf{32} (1977), 185--213; doi:10.1070/RM1977v032n06ABEH003862.

\bibitem{Krichever2} I.M. Krichever, ``Spectral theory of two-dimensional periodic operators and its applications'', \textit{Russian Math. Surveys}, \textbf{44}:2, (1989), 145--225; doi:10.1070/RM1989v044n02ABEH002044.

\bibitem{Krichever3} I.M. Krichever, ``Perturbation Theory in Periodic Problems for Two-Dimensional Integrable Systems'', \textit{Sov. Sci. Rev., Sect. C, Math. Phys. Rev.}, \textbf{9}:2 (1992), 1--103 .

\bibitem{Kuznetsov} E.A. Kuznetsov, ``Solitons in a parametrically unstable plasma'', \textit{Sov. Phys. Dokl.}, \textbf{22} (1977), 507--508.

\bibitem{Yuen2} B.M. Lake, H.C. Yuen, H. Rungaldier, W.E. Ferguson, ``Nonlinear deep-water waves: Theory and experiment. Part 2. Evolution of a continuous wave train'', \textit{J. Fluid Mech.}, \textbf{83}:1 (1977), 49--74; doi:10.1017/S0022112077001037.

\bibitem{Lax} P.D. Lax, ``Periodic solutions of the KdV equation'', \textit{Lectures in Appl. Math.}, \textbf{15} (1974), 85--96.

\bibitem{Lighthill} M.J. Lighthill, Contributions to the theory of waves in nonlinear dispersive systems. J Inst
Math Appl 1:269–306  (1965).

\bibitem{Ma} Y.-C. Ma, ``The perturbed plane wave solutions of the cubic Schr\"odinger equation'', \textit{Stud. Appl. Math.}, \textbf{60}:1 (1979), 43--58; doi:10.1002/sapm197960143.

\bibitem{Malomed}  B. Malomed, "Nonlinear Schrödinger Equations", in Scott, Alwyn (ed.), Encyclopedia of Nonlinear Science, New York: Routledge, pp. 639–643 (2005).

\bibitem{Matveev0} V. B. Matveev and M. A. Salle, {\it Darboux transformations and solitons}, Berlin, Heidelberg: Springer Series in Nonlinear Dynamics, Springer-Verlag, 1991.

\bibitem{MKVM} H.P. McKean, P. Van Moerbeke, ``The spectrum of Hill’s equation'', \textit{Invent. Math.}, \textbf{30}:3 (1975), 217--274; doi:10.1007/BF01425567.

\bibitem{trillo} A. Mussot, C. Naveau, M. Conforti, A. Kudlinski, P. Szriftgiser, F. Copie, S. Trillo, ``Fibre multiwave-mixing combs reveal the broken symmetry of Fermi-Pasta-Ulam recurrence'', \textit{Nature Photonics}, \textbf{12}:5 (2018), 303--308; doi:10.1038/s41566-018-0136-1.

\bibitem{Novikov} S.P. Novikov, ``The periodic problem for the Korteweg-de Vries equation'', \textit{Funct. Anal. Appl.}, \textbf{8}:3 (1974), 236--246; doi:10.1007/BF01075697.

\bibitem{Onorato2} M. Onorato, S. Residori, U. Bortolozzo, A. Montina, F.T. Arecchi, ``Rogue waves and their generating mechanisms in different physical contexts'', \textit{Physics Reports}, \textbf{528}:2 (2013) 47--89; doi:10.1016/j.physrep.2013.03.001.

\bibitem{Osborne} A. Osborne, M. Onorato, M. Serio, ``The nonlinear dynamics of rogue waves and holes in deep-water gravity wave trains'', \textit{Phys. Lett. A}, \textbf{275}(5-6) (2000), 386--393; doi:10.1016/S0375-9601(00)00575-2.

\bibitem{Peregrine} D.H. Peregrine, ``Water waves, nonlinear Schr\"odinger equations and their solutions'', \textit{J. Austral. Math. Soc. Ser. B}, \textbf{25} (1983), 16--43; doi:10.1017/S0334270000003891

\bibitem{PMContiADelRe} D. Pierangeli, F. Di Mei, C. Conti, A.J. Agranat, E. DelRe, ``Spatial Rogue Waves in Photorefractive Ferroelectrics'', \textit{Phys. Rev. Lett.}, \textbf{115}:9 (2015), 093901; doi:10.1103/PhysRevLett.115.093901.

\bibitem{PieranFZMAGSCDR} D. Pierangeli, M. Flammini, L. Zhang, G. Marcucci, A.J. Agranat, P.G. Grinevich, P.M. Santini, C. Conti, E. DelRe, ``Observation of exact Fermi-Pasta-Ulam-Tsingou recurrence and its exact dynamics'', \textit{Phys. Rev. X}, \textbf{8}:4, p. 041017 (9 pages); doi:10.1103/PhysRevX.8.041017;

\bibitem{Pita} L.P. Pitaevskii, S. Stringari, {\it Bose-Einstein Condensation} (Clarendon, Oxford, 2003).

\bibitem{Salasnich} L. Salasnich, A. Parola, L. Reatto, ``Modulational Instability and Complex Dynamics of Confined Matter-Wave Solitons'', \textit{Phys. Rev. Lett.}, \textbf{91}:8, 080405 (2003); doi:10.1103/PhysRevLett.91.080405.

\bibitem{San} P.M. Santini, ``The periodic Cauchy problem for PT-symmetric NLS, I: the first appearance of rogue waves, regular behavior or blow up at finite times'', \textit{J. Phys. A: Math. Theor.}, \textbf{51}:49 (2018), 495207 (21pp); doi:10.1088/1751-8121/aaea05.

  \bibitem{Segur} H. Segur, D. Henderson, J. Carter, J. Hammack, Cong-Ming Li, D. Pheiff, and K. Socha, Stabilizing the Benjamin–Feir instability, J. Fluid Mech. {\bf 539}, 229-271 (2005).

\bibitem{Solli} D.R. Solli, C. Ropers, P. Koonath and B. Jalali, ``Optical rogue waves'', \textit{Nature}, \textbf{450} (2007), 1054--1057; doi:10.1038/nature06402.

\bibitem{Soto} J.M. Soto-Crespo, N. Devine, and N. Akhmediev, Adiabatic transformation of continuous waves into trains of pulses, PHYSICAL REVIEW A 96, 023825 (2017).

\bibitem{Stokes} G. Stokes, ``On the Theory of Oscillatory Waves'', \textit{Transactions of the Cambridge Philosophical Society} \textbf{VIII} (1847) 197--229, and Supplement 314--326.

\bibitem{Taniuti} T. Taniuti and H. Washimi, ``Self-Trapping and Instability of Hydromagnetic Waves along the Magnetic Field in a Cold Plasma'', \textit{Phys. Rev. Lett.}, \textit{21}:4 (1968), 209--212; doi:10.1103/PhysRevLett.21.209.

\bibitem{TBETB} A. Tikan, C. Billet, G. El, A. Tovbis, M. Bertola, T. Sylvestre, F. Gustave, S. Randoux, G. Genty, P. Suret, J. Dudley, ``Universality of the Peregrine soliton in the focusing dynamics of the cubic nonlinear Schr\"odinger equation'', \textit{Phys. Rev. Lett.}, \textbf{119}:3 (2017) 033901; doi:10.1103/PhysRevLett.119.033901.

\bibitem{Tracy} E.R. Tracy, H.H. Chen, ``Nonlinear self-modulation: An exactly solvable model'', \textit{Phys. Rev. A}, \textbf{37}:3 (1988), 815--839; doi:10.1103/PhysRevA.37.815.

\bibitem{Tulin} M.P. Tulin, T. Waseda, ``Laboratory observation of wave group evolution, including breaking effects'', \textit{Journal of Fluid Mechanics}, \textbf{378} (1999), 197--232; doi:10.1017/S0022112098003255.

\bibitem{Simaeys} G. Van Simaeys, P. Emplit, M. Haelterman, ``Experimental Demonstration of the Fermi-Pasta-Ulam Recurrence in a Modulationally Unstable Optical Wave'', \textit{Phys. Rev. Lett.}, \textbf{87}:3 (2001), 033902; doi:10.1103/PhysRevLett.87.033902.

\bibitem{Yuen1} H.C. Yuen, W.E. Ferguson, ``Relationship between Benjamin-Feir instability and recurrence in the nonlinear Schr\"odinger equation'', \textit{Phys. Fluids}, \textbf{21}:8 (1978), 1275--1278; doi:10.1063/1.862394.

\bibitem{Yuen3} H. Yuen, B. Lake, ``Nonlinear dynamics of deep-water gravity waves'', \textit{Advances in Applied Mechanics}, \textbf{22} (1982) 67--229; doi:10.1016/S0065-2156(08)70066-8.

\bibitem{Yurov} A.V. Yurov, and V.A. Yurov, The Landau-Lifshitz Equation, the NLS, and the Magnetic Rogue Wave as a By-Product of Two Colliding Regular ``Positons'', Symmetry, {\bf 10}, 82 (2018). doi:10.3390/sym10040082.

\bibitem{Zakharov} V.E. Zakharov, ``Stability of period waves of finite amplitude on surface of a deep fluid'', \textit{Journal of Applied Mechanics and Technical Physics}, \textbf{9}:2 (1968) 190--194.

\bibitem{ZakharovGelash1} V.E. Zakharov, A.A. Gelash, \textit{Soliton on unstable condensate}, arXiv:1109.0620v2.

\bibitem{ZakharovGelash2} V.E. Zakharov, A.A. Gelash, \textit{On the nonlinear stage of Modulation Instability}, \textit{Phys. Rev. Lett.}, \textbf{111}:5 (2013), 054101; doi:10.1103/PhysRevLett.111.054101.

\bibitem{ZakharovGelash3} V. E. Zakharov, A.A. Gelash, ``Superregular solitonic solutions: a novel scenario for the nonlinear stage of modulation instability'', \textit{Nonlinearity}, \textbf{27}:4 (2014), R1--R39; doi:10.1088/0951-7715/27/4/R1.

\bibitem{ZakharovManakov} V.E. Zakharov, S.V. Manakov, Construction of higher-dimensional nonlinear integrable systems and of their solutions, Functional Analysis and Its Applications, \textbf{19}:2 (1985), 89--101; doi:10.1007/BF01078388.

\bibitem{ZakharovMikha} V. E. Zakharov and A. V. Mikhailov, ``Relativistically invariant two-dimensional models of field theory which are integrable by means of the inverse scattering problem method'', \textit{Sov. Phys. JETP}, \textbf{47} (1978), 1017--1027.

\bibitem{ZakharovOstro} V. Zakharov, L. Ostrovsky, ``Modulation instability: the beginning'', \textit{Physica D: Nonlinear Phenomena}, \textbf{238}:5 (2009), 540--548; doi:10.1016/j.physd.2008.12.002.

\bibitem{ZakharovShabat} V.E. Zakharov, A.B. Shabat, ``Exact theory of two-dimensional self-focusing and one-dimensional self-modulation of waves in nonlinear media'', \textit{Sov. Phys. JETP}, \textbf{34}:1 (1972), 62--69.

\bibitem{ZakharovShabatdress} V.E. Zakharov, A.B. Shabat, ``A scheme for integrating the nonlinear equations of mathematical physics by the method of the inverse scattering transform I'', \textit{Funct. Anal. Appl.} \textbf{8}:3 (1974), 226--235; doi:10.1007/BF01075696.

\end{thebibliography}
\end{document}